\documentclass{aa}

\usepackage{float}
\usepackage{graphicx}
\usepackage{txfonts}
\usepackage{hyperref}
\usepackage{url}
\usepackage{natbib}
\usepackage{multirow}
\begin{document}

   \title{Exploring the crowded central region of 10 Galactic \\ globular clusters using EMCCDs}

   \subtitle{Variable star searches and new discoveries\thanks{based on data collected by the MiNDSTEp team with the Danish 1.54m telescope at ESO’s La Silla observatory in Chile}}

   \titlerunning{Exploring the core of globular clusters with EMCCDs}

   \author{R.~Figuera~Jaimes \inst{1,2} \thanks{email: robertofiguera@gmail.com}
      \and D.~M.~Bramich \inst{3} \thanks{email: dan.bramich@hotmail.co.uk}
      \and J.~Skottfelt  \inst{4,5} 
      \and N.~Kains \inst{6}
      \and U.~G.~J\o{}rgensen \inst{5} 
      \and K.~Horne \inst{1}
	  \and M.~Dominik \inst{1} \thanks{Royal Society University Research Fellow}
	  \and K.~A.~Alsubai \inst{3}
	  \and V.~Bozza \inst{7,8}
	  \and S.~Calchi~Novati \inst{9,7,10} \thanks{Sagan visiting fellow}
	  \and S.~Ciceri \inst{11}
	  \and G.~D'Ago \inst{10}
	  \and P.~Galianni \inst{1}
	  \and S.-H.~Gu  \inst{12,13}
 	  \and K.~B.~W~Harps\o{}e \inst{5}
	  \and T.~Haugb\o{}lle \inst{5}
	  \and T.~C.~Hinse \inst{14}
	  \and M.~Hundertmark \inst{5,1}
	  \and D.~Juncher \inst{5}
	  \and H.~Korhonen \inst{15,5}
	  \and L.~Mancini \inst{11}
	  \and A.~Popovas \inst{5}
	  \and M.~Rabus \inst{16,11}
	  \and S.~Rahvar \inst{17}
	  \and G.~Scarpetta \inst{10,7,8}
	  \and R.~W.~Schmidt \inst{18}
	  \and C.~Snodgrass \inst{19,20}
	  \and J.~Southworth \inst{21}
	  \and D.~Starkey \inst{1}
	  \and R.~A.~Street \inst{22}
	  \and J.~Surdej \inst{23}
	  \and X.-B.~Wang \inst{12,13}
	  \and O.~Wertz \inst{23}
\\
	  (The MiNDSTEp Consortium)
}

   \institute{SUPA, School of Physics and Astronomy, University of St. Andrews, North Haugh, St Andrews, KY16 9SS, United Kingdom.
    \and European Southern Observatory, Karl-Schwarzschild-Stra\ss{}e 2, 85748 Garching bei M\"{u}nchen, Germany 
    \and Qatar Environment and Energy Research Institute (QEERI), HBKU, Qatar Foundation, Doha, Qatar 
    \and Centre for Electronic Imaging, Dept. of Physical Sciences, The Open University, Milton Keynes MK7 6AA, UK 
    \and Niels Bohr Institute and Centre for Star and Planet Formation, University of Copenhagen, {\O}ster Voldgade 5, 1350 Copenhagen K, Denmark 
    \and Space Telescope Science Institute, 3700 San Martin Drive, Baltimore, MD 21218, United States of America 
    \and Dipartimento di Fisica ''E. R. Caianiello'', Universit\`a di Salerno, Via Giovanni Paolo II 132, 84084-Fisciano (SA), Italy 
    \and Istituto Nazionale di Fisica Nucleare, Sezione di Napoli, Napoli, Italy 
    \and NASA Exoplanet Science Institute, MS 100-22, California Institute of Technology, Pasadena CA 91125 
    \and Istituto Internazionale per gli Alti Studi Scientifici (IIASS), 84019 Vietri Sul Mare (SA), Italy 
	\and Max Planck Institute for Astronomy, K\"onigstuhl 17, 69117 Heidelberg, Germany 
	\and Yunnan Observatories, Chinese Academy of Sciences, Kunming 650011, China  
	\and Key Laboratory for the Structure and Evolution of Celestial Objects, Chinese Academy of Sciences, Kunming 650011, China 
	\and Korea Astronomy and Space Science Institute, Daejeon 305-348, Republic of Korea 
	\and Finnish Centre for Astronomy with ESO (FINCA), University of Turku, V{\"a}is{\"a}l{\"a}ntie 20, FI-21500 Piikki{\"o}, Finland 
	\and Instituto de Astrof\'isica, Facultad de F\'isica, Pontificia Universidad Cat\'olica de Chile, Av. Vicu\~na Mackenna 4860, 7820436 Macul, Santiago, Chile 
	\and Department of Physics, Sharif University of Technology, P. O. Box 11155-9161 Tehran, Iran 
	\and Astronomisches Rechen-Institut, Zentrum f\"ur Astronomie der Universit\"at Heidelberg, M\"onchhofstr. 12-14, 69120 Heidelberg, Germany 
    \and Planetary and Space Sciences, Department of Physical Sciences, The Open University, Milton Keynes, MK7 6AA, UK 
    \and Max-Planck-Institute for Solar System Research, Justus-von-Liebig-Weg 3, 37077 G\"ottingen, Germany 
	\and Astrophysics Group, Keele University, Staffordshire, ST5 5BG, UK 
	\and Las Cumbres Observatory Global Telescope Network, 6740 Cortona Drive, Suite 102, Goleta, CA 93117, USA 
	\and Institut d'Astrophysique et de G\'eophysique, Universit\'e de Li\`ege, All\'ee du 6 Ao\^ut, B\^at. B5c, 4000 Li\`ege, Belgium 
    }
   \date{Received September 30, 2015; accepted October 16, 2015}
  \abstract
   {}
   {Obtain time-series photometry of the very crowded central regions of Galactic globular clusters with better angular resolution than previously achieved with conventional CCDs on 
   ground-based telescopes to complete, or improve, the census of the variable star population in those stellar systems.}
   {Images were taken using the Danish 1.54-m Telescope at the ESO observatory at La Silla in Chile. The telescope was equipped with an electron-multiplying CCD and the 
   short-exposure-time images obtained (10 images per second) were stacked using the shift-and-add technique to produce the normal-exposure-time images 
   (minutes). Photometry was performed via difference image analysis. Automatic detection of variable stars in the field was attempted.}
   {The light curves of 12541 stars in the cores of 10 globular clusters were statistically analysed in order to automatically extract the variable stars. We obtained light curves for 31 
   previously known variable stars (3 L, 2 SR, 20 RR Lyrae, 1 SX Phe, 3 cataclysmic variables, 1 EW and 1 NC) and we discovered 30 new variables (16 L, 7 SR, 4 RR Lyrae, 1 SX Phe and 2 NC).}
   {}
   \maketitle
%
\section{Introduction}

Globular cluster systems in the Milky Way are excellent laboratories for several fields in astronomy, from cosmology 
to stellar evolution. As old as our Galaxy, they are stellar fossils that enable astronomers to look back to earlier galaxy 
formation stages to have a better understanding of the stellar population that is forming the clusters.

Possibly the first official register of globular clusters started with the ``Catalog of Nebulae and Star Clusters'' made by \cite{Messier1781} and followed by \cite{herschel1786}. Several early studies \citep[i. e. ][]{barnard1909, Hertzsprung1915, Eddington1916, shapley1916, oort1924} of globular clusters can also be found in the literature.

The first variable star found in the field of a globular cluster was T Scorpii \citep{luther1860}. Subsequent early studies of variable stars in globular clusters can also be found in the literature \citep[see e. g.][]{davis1917, bailey1918, sawyer1931, oosterhoff1938}. Measurements of the variability of these stars were done either by eye or by using photographic plates. It was not until about 1980 \citep{GCschool99} that deeper studies in the photometry of globular clusters were done thanks to both the use of CCDs and larger telescopes.

More recently, the implementation of new image analysis tools such as difference image analysis \citep[DIA;][]{alard98+01}, have enabled much more detailed and quantitative studies, but there are still several limiting constraints. The first is saturated stars. It is always important to balance the exposure time to prevent bright stars from saturating the CCD without losing signal-to-noise ratio in the fainter stars. The second difficulty is the crowded central regions. Sometimes it is complicated to measure properly the point spread function (PSF) of the stars in the central regions of globular clusters due to blending caused by the effect of atmospheric turbulence.

To overcome these limitations, we started a pilot study of the crowded central regions of globular clusters using electron-multiplying CCDs 
(EMCCDs) and the shift-and-add technique. Our globular cluster sample is chosen based mainly on the central concentration of stars while favouring 
clusters well known to have variable stars. We published the results of globular cluster NGC~6981 in \cite{skottfelt13} with the discovery of two new variables and five more globular clusters in \cite{Skottfelt15+05} where a total of 114 previously 
unknown variables were found. In this paper we report the results on 10 further globular clusters.

Sec. \ref{sec:data_red} describes the instrumentation, data obtained, pipeline used and the data reduction procedures. Secs. \ref{sec:calib}, \ref{sec:cmd} and \ref{sec:var_search} explain the calibration, the colour magnitude diagrams, and the strategy used to identify variable 
stars. In Sec. \ref{sec:var_class} we classify the variable stars. Sec. \ref{sec:cluster_obs} describes the methodology followed to observe the globular clusters. Sec. \ref{sec:results} presents the results obtained in each globular cluster and Sec. \ref{sec:conclusion} summarises the main conclusions.

\section{Data and reductions}\label{sec:data_red}

\subsection{Telescope and instrument}\label{subsec:instrument}

Observations were taken using the 1.54 m Danish telescope at the ESO observatory at La Silla in Chile, which is located at an elevation of 2340 m at 70$^{\circ}$44$^ 
{\prime}$07$^{\prime\prime}$.662W 29$^{\circ}$15$^{\prime}$14$^{\prime\prime}$.235S. The telescope is equipped with an Andor Technology iXon+897 EMCCD camera, which has a 512 $\times$ 512 array of 16 $\mu$m pixels, a pixel scale of 0$^{\prime\prime}$.09 per pixel and a total field of view of $\sim45\times45$ arcsec$^{2}$.

For the purpose of this research the camera was configured to work at a frame-rate of 10 Hz (this is 10 images per second) and an EM gain of $300e^-/$photon. The camera is placed behind a dichroic mirror which works as a long-pass filter.
Considering the mirror and the sensitivity of the camera it is possible to cover a wavelength range between 650 nm to 1050 nm, corresponding roughly to a combination of SDSS $\mathrm{i}^{\prime}+\mathrm{z}^{\prime}$ filters \citep{bessell05}. More details about the instrument can be found in \cite{Skottfelt15+15}.

\subsection{Observations}

\begin{figure*}[htp!]
\centering
\includegraphics[scale=1]{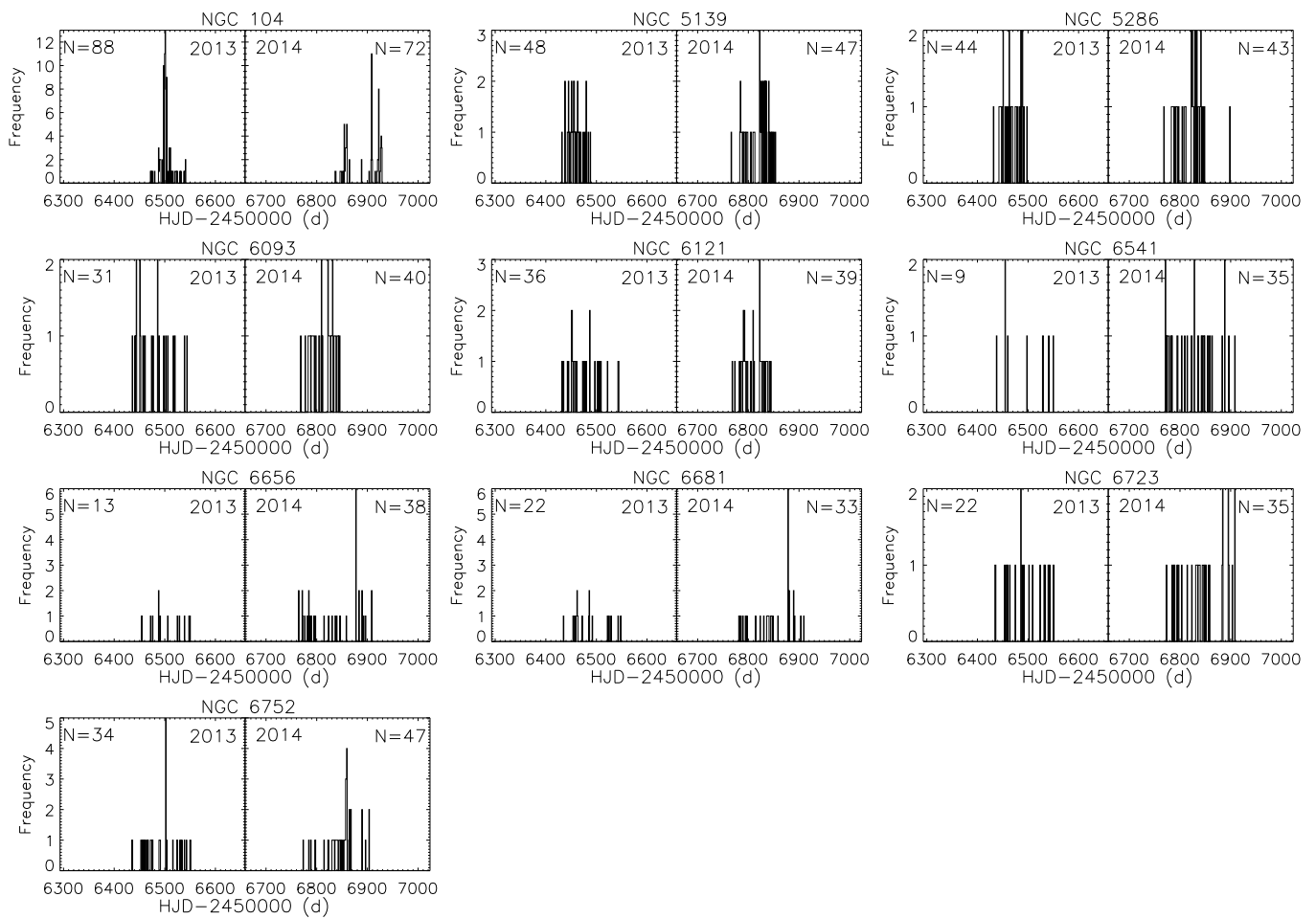}
\caption{Histograms with the number of observations per night for each globular cluster studied in this work. In the panels on the left is plotted data taken during 2013 and in the panels on the right, data taken during 2014.}
\label{fig:histograms_nights}
\end{figure*}

Observations were taken during 2013 and 2014 as part of an ongoing program at the Danish telescope that was implemented from April to September each year. Fig. \ref{fig:histograms_nights} shows histograms of the number of observations per night per cluster. Data in the left panel corresponds to 2013 and data in the right panel, to 2014. As the time at the telescope was limited, we tried to observe all clusters with the same 
rate as far as it was possible by taking two observations each night. However, as it is shown in the histograms, it was not possible due to weather conditions or because the telescope was used to monitor microlensing events carried out by the MINDSTEp consortium as part of the program to characterise exoplanets.

It is worth commenting that, depending on the magnitude level of the horizontal branch of each globular cluster, observations with total exposure times between 6 to 10 minutes were produced (see Sec. \ref{sec:cluster_obs}). To do this the camera was continuously taking images (at the rate of 10 images per second) for the total exposure time desired and then all the images obtained during that particular exposure time were stacked to produce a single observation. That is, a 10 minutes exposure time observation is the result of stacking 6000 images that were continuously taken. The technique for image stacking is called ``shift-and-add'' and it is explained in Sec. \ref{sec:stacked_image}.

\subsection{EMCCD data reduction}

EMCCDs also known as low light level charge-coupled devices (L3CCD) \citep[see e. g.][]{smith08+06, jerram01+09} are conventional CCDs that are equipped with an extended serial gain register in 
which each electron produced during the exposure time has a probability to create a second electron (avalanche multiplication) when it is passing 
through the extended gain register. The process in which the second electrons are created is called impact ionisation.

To do bias, flat-field and tip-tilt corrections, the procedures and algorithms described in \cite{harpsoe12+03} were used.

\subsubsection{Tip-tilt correction}

As described in \cite{harpsoe12+03}, the tip-tilt correction for a sample of bias and flat-field corrected images $I_i(x, y)$ with respect to a reference 
image $R(x, y)$ is calculated by finding the global maximum of $P_i$ defined by Eqn. \ref{eq:tip_tilt},

\begin{equation}\label{eq:tip_tilt}
P_i(x,y)=|FFT^{-1}[FFT(R)\cdot\overline{FFT(I_i)}]|,
\end{equation}
which is the cross-correlation image (see Fourier cross correlation theorem). The position of the maximum ($x_{\mathrm{max}}, y_{\mathrm{max}}$) is used to shift each of the images in the sample $I_i(x, y)$ to the position of the stars in the reference image.

\subsubsection{Image quality and stacking}\label{sec:stacked_image}

The image quality is assessed by defining the quality index
\begin{equation}\label{eq:image_quality}
q_i = \frac{P_i(x_{\text{max}},y_{\text{max}})}{\sum_{\substack{\left|(x-x_{\text{max}},y-y_{\text{max}})\right| < r \\ (x,y)\neq(x_{\text{max}},y_{\text{max}})}} P_i(x,y)},
\end{equation}
which represents the ratio of the maximum $P_i(x_{\mathrm{max}},y_{\mathrm{max}})$ to the sum of the $P_i(x, y)$ values of the surrounding pixels. Once this index is calculated, images from a single observation are sorted from the 
one with the best quality (smallest $q_i$) to the one with the worst quality (largest $q_i$) and a ten-layer image frame is built by stacking images in the sequence: 1\%, 2\%, 5\%, 10\%, 20\%, 50\%, 90\%, 98\%, 99\% and 100\%. This means that the first-layer frame has stacked the top 1\% of the best quality images, the second-layer frame has stacked the next 1\% of the best quality images, the third-layer frame has 
stacked the next 3\%, the fourth has stacked the next 5\% and so on. In this study, we use the layers with the best FWHM over all the frames to build the reference image for each globular cluster and we stack the ten-layers of an observation into a single science image for the photometry \citep[see also][]{Skottfelt15+05}.

\subsection{Photometry}

To extract the photometry in each of the images we used the DanDIA\footnote{DanDIA is built from the DanIDL library 
of IDL routines available at \url{http://www.danidl.co.uk}} pipeline \citep{bramich08, bramich13+10}, which is based on difference image analysis (DIA).

As commented in Sec. \ref{sec:stacked_image}, for each EMCCD observation, we produced a ten-layer calibrated image cube where each of the images in the cube are sorted by quality. DanDIA 
builds a high $S/N$ and high-resolution reference frame by selecting and combining the best quality images available in the cubes. This is done by detecting 
bright stars using DAOFIND \citep{stetson87} and employing these to align the images using the triangulation technique described in \cite{pal06+01}. Each of the reference frames used in the analysis of each globular cluster 
can be found throughout the paper (e. g. Fig. \ref{fig:finding_chart_NGC104}) and the mean PSF FWHM for each reference image is listed in Tab. \ref{tab:clusters}. Positions and reference 
fluxes ($f_{\mathrm{ref}}$) in ADU/s for each star are calculated using the PSF photometry package "STARFINDER" \citep{diolaiti00+05}. The pipeline implements a low-order polynomial 
degree for the spatial variation of the PSF model (a quadratic polynomial degree was sufficient in our case).

Once the reference image is built, all of the science frames are registered with the reference image by using again the \cite {stetson87} and \cite{pal06+01} algorithms described before. Image subtraction 
then determines a spatially variable kernel, modelled as a discrete pixel array, that best matches the reference image to each science image. The photometric scale factor ($p(t)$) used to scale the reference frame to each image is calculated as part of the kernel model \citep{bramich15+04}. Difference images are created by subtracting the convolved reference image from each science image. Finally, difference fluxes ($f_{\mathrm{diff}}(t)$) in ADU/s for each star detected in the reference image are measured in each of the 
difference images by optimally scaling the point spread function model for the star to the difference image. The light curve for 
each star is the total flux ($f_{\mathrm{tot}}(t)$) in ADU/s defined as
\begin{equation}\label{eq:total_flux}
f_{\mathrm{tot}}(t)=f_{\mathrm{ref}}+\frac{f_{\mathrm{diff}}(t)}{p(t)}.
\end{equation}

We transform this to instrumental magnitudes $m_{\mathrm{ins}}$ at each given time $t$:
\begin{equation}\label{eq:inst_mag}
m_{\mathrm{ins}}(t)=17.5-2.5\log(f_{\mathrm{tot}}(t)).
\end{equation}

A detailed description of the procedures and techniques employed by the pipeline can be found in \cite{bramich11+03}.

For a conventional CCD, the noise model used by DanDIA is represented by the pixel variances $\sigma_{kij}^2$ of the image $k$ at the pixel positions $i$, $j$
\begin{equation}\label{variance_ccd}
\sigma_{kij}^2=\frac{\sigma_0^2}{F_{ij}^2}+\frac{M_{kij}}{GF_{ij}},
\end{equation}
where $\sigma_0$ is the CCD readout noise (ADU), $F_{ij}$ is the master flat-field image, $G$ is the CCD gain ($e^-$/ADU) and $M_{kij}$ is the image model. For an EMCCD, the noise model for a single exposure is different:
\begin{equation}\label{variance_emccd}
\sigma_{ij}^2=\frac{\sigma_0^2}{F_{ij}^2}+\frac{2\,M_{ij}}{F_{ij}\,G_{\mathrm{EM}}\,G_{\mathrm{PA}}},
\end{equation}
where $G_{\mathrm{EM}}$ is the electron-multiplying gain (photons/$e^-$) and $G_{\mathrm{PA}}$ is the Pre Amp gain ($e^-$/ADU).

If $N$ exposures are combined by summation, then the noise model for the combined image $\sigma_{ij,\mathrm{comb}}^{2}$ is given by Eqn. \ref{variance_emccd_combined}:
\begin{equation}\label{variance_emccd_combined}
\sigma_{ij,\mathrm{comb}}^2=\frac{N\,\sigma_0^2}{F_{ij}^2}+\frac{2\, M_{ij,\mathrm{comb}}}{F_{ij}\,G_{\mathrm{EM}}\,G_{\mathrm{PA}}},
\end{equation}
where $M_{ij, \mathrm{comb}}$ is the image model of the combined image \citep[see also][]{Skottfelt15+05}.

In Tab. \ref{tab:electronic_data}, we illustrate the format of the electronic table with all fluxes and photometric measurements as they are available at the CDS\footnote{http://cds.u-strasbg.fr/}.

\begin{table*}[htp!]
\caption{Time-series I photometry for all known and new variables in the field of view covered in each globular cluster.
The standard $M_{\mathrm{std}}$ and instrumental $M_{\mathrm{ins}}$ magnitudes are listed in columns 5 and 6, respectively, corresponding to the cluster, variable star, filter, and epoch of mid-exposure
listed in columns 1-4, respectively. The uncertainty on $M_{\mathrm{ins}}$ is listed in column 7, which also
corresponds to the uncertainty on $M_{\mathrm{std}}$. For completeness, we also list the quantities $f_{\mathrm{ref}}$, $f_{\mathrm{diff}}$ and $p$ from Eq. \ref{eq:total_flux} in columns 8, 10 and 12, along with
the uncertainties $\sigma_{\mathrm{ref}}$ and $\sigma_{\mathrm{diff}}$ in columns 9 and 11. This is an extract from the full table, which is available with the electronic version of the article at the CDS.}
\label{tab:electronic_data}
\centering
\tabcolsep=0.08cm
\begin{tabular}{cccccccccccccc}
\hline\hline
Cluster & var & Filter & HJD & $M_{\mathrm{std}}$ & $M_{\mathrm{ins}}$ & $\sigma_m$ & $f_{\mathrm{ref}}$ & $\sigma_{\mathrm{ref}}$ & $f_{\mathrm{diff}}$ & $\sigma_{\mathrm{diff}}$ & $p$ \\
        & id  &        & (d) &       (mag)        &       (mag)        &   (mag)    & (ADU s$^{-1}$)    &     (ADU s$^{-1}$)     &   (ADU s$^{-1}$)   & (ADU s$^{-1}$) &     \\
\hline
NGC104  & PC1-V12 & I      & 2456472.93210 & 15.846 & 6.777 & 0.014 & 24971.015 & 9644.446 & -30350.989 & 1425.567 & 5.5131\\
NGC104  & PC1-V12 & I      & 2456476.94184 & 15.729 & 6.659 & 0.012 & 24971.015 & 9644.446 & -16234.626 & 1138.337 & 4.9447\\
\vdots  & \vdots  & \vdots &      \vdots   & \vdots &\vdots &\vdots &\vdots     &  \vdots     &   \vdots     &  \vdots  & \vdots\\
NGC5139 &   V457  & I      & 2456432.58154 & 15.735 & 6.649 & 0.007 & 20096.379 & 684.928 & +8409.333 & 673.724 & 4.6618\\
NGC5139 &   V457  & I      & 2456438.70588 & 15.832 & 6.746 & 0.005 & 20096.379 & 684.928 & -312.498 & 432.554 & 4.5021\\
\vdots  & \vdots  & \vdots &      \vdots   & \vdots &\vdots &\vdots &\vdots     &  \vdots     &   \vdots     &  \vdots  & \vdots\\
NGC5286 &    V37  & I      & 2456445.53773 & 15.978 & 6.951 & 0.005 & 18451.916 & 1202.455 & -8631.145 & 385.937 & 4.6006\\
NGC5286 &    V37  & I      & 2456446.53053 & 15.744 & 6.717 & 0.003 & 18451.916 & 1202.455 & +9806.700 & 232.129 & 4.6341\\
\vdots  & \vdots  & \vdots &      \vdots   & \vdots &\vdots &\vdots &\vdots     &  \vdots     &   \vdots     &  \vdots  & \vdots\\

\hline
\end{tabular}
\end{table*}

\subsection{Astrometry and finding chart}\label{ssec:astro_chart}

To fit an astrometric model for the reference images, celestial coordinates available at the ``ACS Globular Cluster Survey''\footnote{http://www.astro.ufl.edu/\textasciitilde ata/public\_hstgc/} \citep[see][]{anderson08+12} were uploaded for the field of the cluster through GAIA (Graphical Astronomy and Image Analysis Tool; \cite{gaia}). An ($x$,$y$) shift was applied to all of the uploaded positions until they matched the stars in the fields. Stars lying outside the field of view and those without a clear 
match were removed, and the (x, y) shift was refined by minimising the squared coordinate residuals. The number of stars used in the matching process ranged from 108 stars to 1034 stars which guaranteed that the astrometric solutions applied to the reference images considered enough stars covering the whole field. The 
radial RMS scatter obtained in the residuals was $\sim$ 0$^{\prime\prime}$.1 ($\sim$ 1 pixel). The astrometrically calibrated reference images were used to produce a finding chart for each globular cluster on which we marked the positions and identifications of all variable stars studied in this work. Finally, a table with the equatorial J2000 celestial coordinates of all variables for each globular cluster is given (see e. g. Tab. \ref{tab:NGC104_ephemerides}).

\section{Photometric calibration}\label{sec:calib}

The photometric transformation of instrumental magnitudes to the standard system was accomplished using information available in the ACS Globular 
Cluster Survey, which provides calibrated magnitudes for selected stars in the fields of 50 globular clusters extracted from images taken with the Hubble Space 
Telescope (HST) instruments ACS and WFPC.

\begin{figure*}[htp!]
\centering
\includegraphics[scale=1]{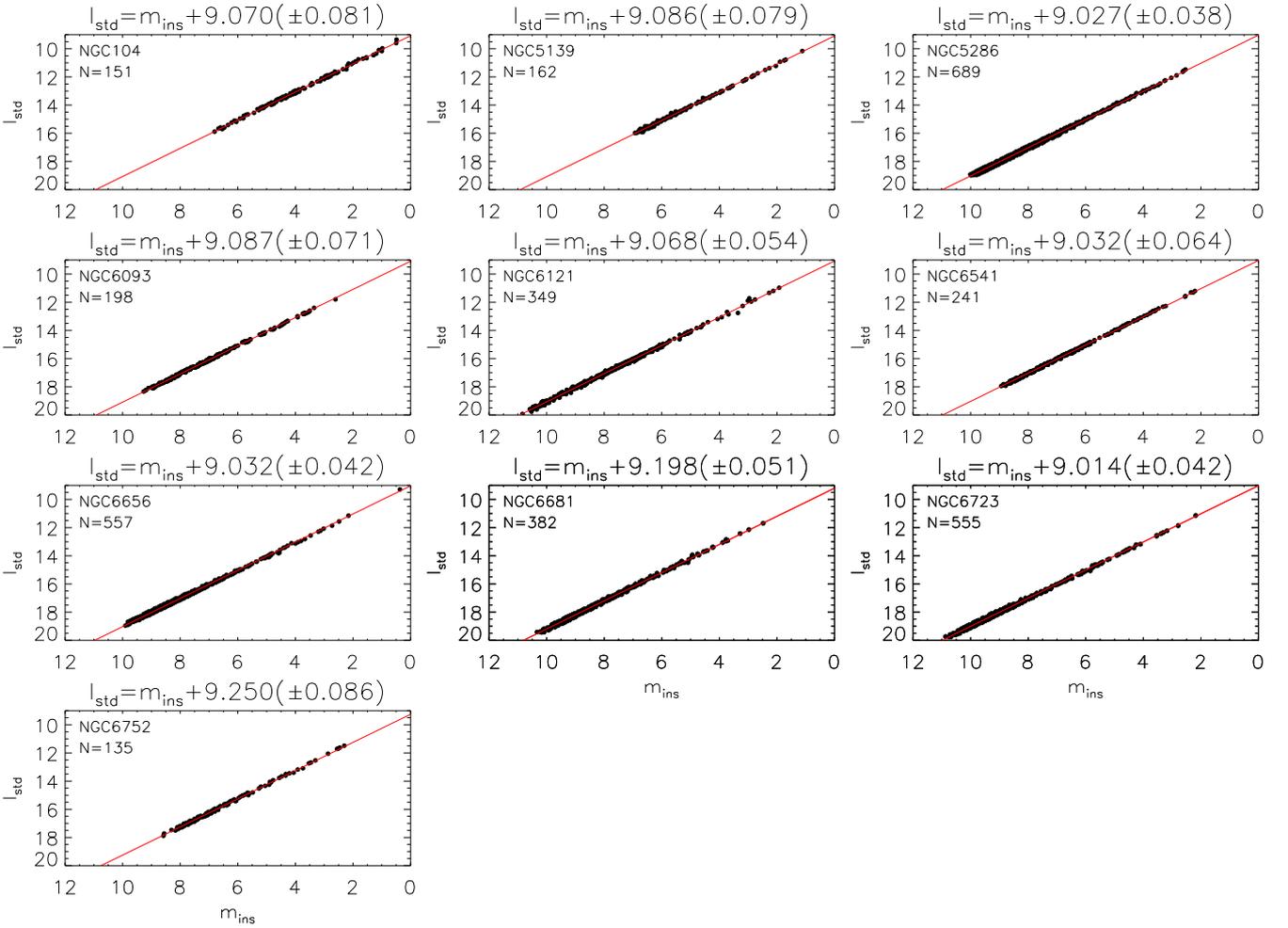}
\caption{Standard I magnitude taken from the HST observations as a function of the instrumental $\mathrm{i}^{\prime}+\mathrm{z}^{\prime}$ magnitude. The red lines are the fits that 
best match the data and they are described by the equations in the titles of each plot. The correlation coefficient is 0.999 in all cases.}
\label{fig:std_calibration}
\end{figure*}

By matching the positions of the stars in the field of the HST images with those in our reference images, we obtained photometric transformations for our 10 globular clusters, as shown 
in Fig. \ref{fig:std_calibration}. The I magnitude obtained from the ACS \citep[see][]{sirianni05+13} is plotted versus the instrumental $\mathrm{i}^{\prime}+\mathrm{z}^{\prime}$ magnitude obtained in 
this study. The red lines are linear fits with slope unity yielding zero-points labelled in the title of each plot. $N$ is the number of stars used in the fit. The correlation 
coefficient was 0.999 in all cases. A transformation was derived for each globular cluster.

\section{Colour magnitude diagrams}\label{sec:cmd}

As our sample has data available for only one filter, we decided to build colour magnitude diagrams (CMD) (see Fig. \ref{fig:cmd_NGC104}) by 
using the information available from the HST images at the ACS Globular Cluster Survey. The data used correspond to the V and I photometry 
obtained in \cite{sirianni05+13}. The CMD was useful in classifying the variable stars, especially those with poorly defined light curves such as long-period irregular variables and semi-regular variables, as well as corroborating cluster membership.

\section{Variable star searches}\label{sec:var_search}

During the reductions, the DanDIA pipeline produced a total of 12541 light curves for stars in the cores of the 10 globular clusters studied. We used three automatic 
(or semi-automatic) techniques to search for variable stars, as described in Secs. \ref{sec:rms}, \ref{sec:sb} and \ref{sec:sum_diff}.

\subsection{Root mean square}\label{sec:rms}

Diagrams of root mean square (RMS) magnitude deviation against mean I magnitude (see top Fig. \ref{fig:rms_sb_NGC0104}) were 
constructed for each globular cluster. In these diagrams we measure the photometric scatter for each 
star studied in this work, but also we measure the intrinsic variation of the variable stars over time, which gives them a higher RMS than the none variables. The classification is indicated 
by the colour as detailed in Tab. \ref{tab:var_type}.

To select possible variable stars, we fit a polynomial model to the RMS values as a function of magnitude and flag all stars with RMS values greater than 3 times the model. All 
difference images obtained with DanDIA were blinked as well to corroborate the variation of the stars selected.

\subsection{$S_B$ statistic}\label{sec:sb}

A detailed discussion can be found about the benefits of using the $S_B$ statistic to detect variable stars \citep{figuera13+04} and 
RR Lyrae with Blazhko effect \citep{arellano12+04}. The $S_B$ statistic is defined as 
\begin{equation}\label{eq:sb_index}
S_B=\left(\frac{1}{NM}\right)\sum_{i=1}^M\left(\frac{r_{i,1}}{\sigma_{i,1}}+\frac{r_{i,2}}{\sigma_{i,2}}+...+\frac{r_{i,k_i}}{\sigma_{i,k_i}}\right)^2,
\end{equation}
where $N$ is the number of data points for a given light curve and $M$ is the number of groups formed of time-consecutive residuals of the same sign from a 
constant-brightness light curve model (e. g. mean or median). The residuals $r_{i,1}$ to $r_{i,k_i}$ correspond to the $i$th group of $k_i$ time-consecutive residuals of 
the same sign with corresponding uncertainties $\sigma_{i,1}$ to $\sigma_{i,k_i}$. The $S_B$ statistic is larger in value for light curves with 
long runs of consecutive data points above or below the mean, which is the case for variable stars with periods longer than the typical photometric cadence.

Plots of $S_B$ versus mean I magnitude are given for each globular cluster (see bottom Fig. \ref{fig:rms_sb_NGC0104}) and variable stars are plotted in colour.

To select variable stars, the same technique employed in \ref{sec:rms} was used, but in this case the threshold was shifted between 3-10 times the model $S_B$ values depending on the distribution of the data in each globular cluster. All stars selected were inspected in the difference images in the same way as explained in Sec. \ref{sec:rms}.

\subsection{Stacked Difference images}\label{sec:sum_diff}

Based on the results obtained using the DanDIA pipeline, a stacked difference image was built for each globular cluster with the aim of detecting the difference fluxes that correspond to variable stars in the field of the reference image. The stacked image is the result of summing the absolute values of the difference images divided by the respective pixel uncertainty
\begin{equation}\label{eq:sum_diff}
S_{ij}=\sum_k\frac{|D_{kij}|}{\sigma_{kij}},
\end{equation}
where $S_{ij}$ is the stacked image, $D_{kij}$ is the $k$th difference image, $\sigma_{kij}$ is the pixel uncertainty associated with each image $k$ and the indexes $i$ and $j$ correspond to pixel positions.

All of the variable star candidates obtained by using the RMS diagrams or the $S_B$ statistic explained in Secs. \ref{sec:rms} and \ref{sec:sb} were inspected visually in the stacked images to confirm or refute their variability. Difference images were also blinked to finally corroborate the variation of the stars selected.

\section{Variable star classification}\label{sec:var_class}

All variable stars found were plotted in the colour-magnitude diagram for the corresponding globular cluster, to have a better understanding of the nature of their variation and their cluster membership. A period search of the light curve variations was undertaken by using the string method \citep{SQmethod} and by minimising the $\chi^2$ in a Fourier analysis 
fit. To decide the classification of the variable stars, we used the conventions defined in the General Catalogue of Variable Stars \citep{gcvs} and by considering their position in the colour-magnitude diagrams and the periods found.

In Tab. \ref{tab:var_type}, the classification, corresponding symbols and colours used in the plots throughout the paper are shown. 

\begin{table}[htp!]
\caption{Convention used in the variable star classification of this work based on the definitions of the general catalogue of variable stars \citep{gcvs}.}
\label{tab:var_type}
\centering
\tabcolsep=0.08cm
\small{
\begin{tabular}{cccc}
\hline\hline
Type & Id & Point style & Color \\
\hline
&\multicolumn{3}{l}{\hspace{0.5cm}\textbf{Pulsating Variables}} \\
\hline
RR Lyrae (RRL)  & RR0    & Filled circle               & Red     \\
                & RR01   & Filled circle               & Blue    \\
                & RR1    & Filled circle               & Green   \\
Semi-regular    & SR     & Filled square               & Red     \\
Long-period irregular    & L      & Filled square               & Purple  \\
SX Phoenicis    & SX Phe & Filled triangle             & Cyan    \\
\hline
&\multicolumn{3}{l}{\hspace{0.5cm}\textbf{Cataclysmic Variables}} \\
\hline
In general      & \textbf{DN, Novae} & Filled five pointed star    & Green   \\
\hline
&\multicolumn{3}{l}{\hspace{0.5cm}\textbf{Eclipsing Variables}} \\
\hline
                & EW     & Filled five pointed star       & Blue    \\
\hline
&\multicolumn{3}{l}{\hspace{0.5cm}\textbf{Unclassified Variables}} \\
\hline
In general  & NC     & Filled square               & Yellow  \\

\hline
\end{tabular}}
\end{table}

\section{Globular clusters observed}\label{sec:cluster_obs}

\begin{figure}[htp!]
\centering
\includegraphics[scale=1.25]{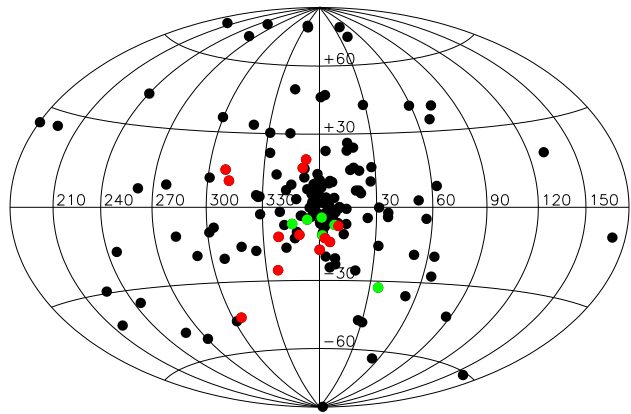}
\caption{Positions of the 157 known globular clusters in the Galaxy are plotted in the Galactic plane system. The 10 globular clusters studied in this work are labelled in red. Green corresponds to the globular clusters studied in \cite{Skottfelt15+05, skottfelt13}.}
\label{fig:galactic_map}
\end{figure}

Fig. \ref{fig:galactic_map}, shows the Galactic plane celestial coordinates (l, b) for the 157 globular clusters listed in \cite[][2010 edition]{harriscatalog}. The 10 globular clusters analysed in this work are plotted in red colour where 6 are in the Southern hemisphere and 4 are in the Northern hemisphere. Also plotted (in green) are the 6 globular clusters studied in \cite{Skottfelt15+05, skottfelt13}.

The strategy to select and observe the clusters was the following:

- (1) We take into account the visibility. We selected all globular clusters visible based on the geographical coordinates of the telescope and some pointing restrictions too. 

- (2) We focused on the concentration of stars in the central region of the clusters. We used as reference the central luminosity density $\rho_0$ ($L_{\odot}$pc$^{-3}$) listed in \cite[][2010 edition]{harriscatalog}. The entire sample available in the catalogue has $\log(\rho_0)$ from $\sim$0 to $\sim$6. To 
explore the potential of the EMCCD and the shift-and-add technique for photometry, a sample with different levels of $\rho_0$ were chosen from a very dense cluster
NGC~6752 ($\log(\rho_0)$=5.04) to a less dense cluster NGC~6723 ($\log(\rho_0)$=2.79) and some intermediate levels as NGC~5139 ($\log(\rho_0)$=3.15) and NGC~5286 ($\log(\rho_0)$=4.10).

- (3) A bibliography review revealed how many variable stars are known from previous work. To do this we used the information available for each globular cluster in the Catalogue of Variable Stars in Galactic Globular 
Clusters \citep{clement01} and the ADS to do the bibliography review up-to-now.

- (4) Exploration of known colour-magnitude diagrams. As most of the variable stars that are globular cluster members have a particular position in these diagrams, we 
focused our attention on globular clusters with a high concentration of stars for example at the top of the red giant branch (for semi-regular variables), the instability strip of the horizontal branch (for RR Lyrae) and the 
blue straggler region (for SX Phoenicis). To do this we used the colour information available in the ACS to build colour-magnitude diagrams for each globular cluster. Additionally we 
also used the Galactic globular clusters database\footnote{http://gclusters.altervista.org/}.

- (5) The exposure time to be used for a single EMCCD observation for each globular cluster was chosen based on the V magnitude of the horizontal branch. We selected exposure times as follows:

\begin{center}

6 min for $V_{\mathrm{HB}}$ < 14 mag,

8 min for 14 mag < $V_{\mathrm{HB}}$ < 17 mag,

10 min for $V_{\mathrm{HB}}$ > 17 mag.

\end{center}

We chose the 10 globular clusters presented in this work. Some of their most relevant physical properties are detailed in 
Tab. \ref{tab:clusters}.

\begin{table*}[htp!]
\caption{Some of the physical properties of the globular clusters studied in this work. Column 1 is the section with the individual results for each globular cluster. Column 2, is the name of the cluster as it is defined in the New General Catalogue, Columns 3 and 4 are the 
celestial coordinates (right ascension and declination), Column 5 is the distance from the Sun, Column 6 is distance from the Galactic centre, Column 7 is metallicity, Column 8 is reddening, Column 9 is V magnitude level of the horizontal branch, Column 10 is V distance modulus, Column 11 is King-model central 
concentration c = $\log$(r$_t$/r$_c$), Column 12 is central luminosity density $\log_{10}(L_{\odot}pc^{-3})$, Column 13 is the mean full-width half-maximum (arcsec) measured in the reference image and Column 14 is the exposure time in the reference image.}
\label{tab:clusters}
\centering
\tabcolsep=0.13cm
\begin{tabular}{cccccccccccccc}
\hline\hline
Section &Cluster & RA    & Dec   & $D_{\odot}$ & $D_{gc}$ & [Fe/H] & E(B-V) & V$_{\mathrm{HB}}$ & $V-M_V$ & c &   $\rho_0$       & FWHM   & $t_{\mathrm{exp}}$\\
        &  NGC   & J2000 & J2000 &   kpc       &  kpc     &        & mag    &        mag        &  mag    &   &$log(L_{\odot}\mathrm{pc}^{-3})$& arcsec & s \\
\hline
\ref{sec:NGC104}  & 104  & 00:24:05.67 & -72:04:52.6 & 4.5  & 7.4 & -0.72 & 0.04 & 14.06 & 13.37 & 2.07 & 4.88 & 0.61 & 302.4\\
\ref{sec:NGC5139} & 5139 & 13:26:47.24 & -47:28:46.5 & 5.2  & 6.4 & -1.53 & 0.12 & 14.51 & 13.94 & 1.31 & 3.15 & 0.38 & 312.0\\
\ref{sec:NGC5286} & 5286 & 13:46:26.81 & -51:22:27.3 & 11.7 & 8.9 & -1.69 & 0.24 & 16.63 & 16.08 & 1.41 & 4.10 & 0.37 & 316.8\\
\ref{sec:NGC6093} & 6093 & 16:17:02.41 & -22:58:33.9 & 10.0 & 3.8 & -1.75 & 0.18 & 16.10 & 15.56 & 1.68 & 4.79 & 0.45 & 302.4\\
\ref{sec:NGC6121} & 6121 & 16:23:35.22 & -26:31:32.7 & 2.2  & 5.9 & -1.16 & 0.35 & 13.45 & 12.82 & 1.65 & 3.64 & 0.49 & 180.0\\
\ref{sec:NGC6541} & 6541 & 18:08:02.36 & -43:42:53.6 & 7.5  & 2.1 & -1.81 & 0.14 & 15.35 & 14.82 & 1.86 & 4.65 & 0.56 & 340.8\\
\ref{sec:NGC6656} & 6656 & 18:36:23.94 & -23:54:17.1 & 3.2  & 4.9 & -1.70 & 0.34 & 14.15 & 13.60 & 1.38 & 3.63 & 0.50 & 297.6\\
\ref{sec:NGC6681} & 6681 & 18:43:12.76 & -32:17:31.6 & 9.0  & 2.2 & -1.62 & 0.07 & 15.55 & 14.99 & 2.50 & 5.82 & 0.49 & 312.0\\
\ref{sec:NGC6723} & 6723 & 18:59:33.15 & -36:37:56.1 & 8.7  & 2.6 & -1.10 & 0.05 & 15.48 & 14.84 & 1.11 & 2.79 & 0.51 & 360.0\\
\ref{sec:NGC6752} & 6752 & 19:10:52.11 & -59:59:04.4 & 4.0  & 5.2 & -1.54 & 0.04 & 13.70 & 13.13 & 2.50 & 5.04 & 0.59 & 198.0\\

\hline
\end{tabular}
\end{table*}

\section{Results}\label{sec:results}

\subsection{\textbf{NGC~104 / C0021-723 / 47 Tucanae}}\label{sec:NGC104}

This globular cluster was discovered by Nicholas Louis de Lacaille in 1751\footnote{http://messier.seds.org/xtra/ngc/n0104.html}. The 
cluster is in the constellation of Tucana at a distance of 4.5 kpc from the Sun and 7.4 kpc from the Galactic centre. It has a 
metallicity of [Fe/H]=$-$0.72 dex and a distance modulus of (m$-$M)$_{V}$=13.37 mag. The magnitude of its horizontal branch is 
V$_{\mathrm{HB}}$=14.06 mag.

\begin{table*}[htp!]
\caption{NGC~104: Ephemerides and main characteristics of the variable stars in the field of this globular cluster. Column 1 is the id assigned to the variable star, Columns 2 and 3 
correspond to the right ascension and declination (J2000), Column 4 is the epoch used, Column 5 is the period, Column 6 is median of the data, Column 7 is the peak-to-peak amplitude in the light curve, Column 8 is the number of epochs and Column 9 is the classification of the variable. The numbers in parentheses indicate the uncertainty on the last decimal place of the period.}
\label{tab:NGC104_ephemerides}
\centering
\begin{tabular}{ccccccccc}
\hline\hline
Var id &   RA  &  Dec  &Epoch & $P$ & $I_{median}$ & $A_{\mathrm{i}^{\prime}+\mathrm{z}^{\prime}}$ & $N$ & Type\\
       & J2000 & J2000 & HJD  & d &      mag     &                mag                            &   &     \\
\hline
PC1-V12 & 00:24:05.921 & -72:04:45.20 &      --      &    --        & 15.62  & 1.10 & 149 & NC \\
WF2-V34 & 00:24:08.406 & -72:04:35.91 &      --      &    --        & 10.98  & 0.03 & 139 & L \\
LW10    & 00:24:02.490 & -72:05:07.45 &      --      &    --        &  9.60  & 0.24 & 108 & L \\
LW11    & 00:24:03.145 & -72:04:50.60 & 2456920.7253 & 19.37(8)     & 10.18  & 0.08 & 141 & SR \\
LW12    & 00:24:03.982 & -72:05:10.06 &      --      &    --        &  9.57  & 0.09 &  41 & L \\
\hline
EM1  & 00:24:03.065 & -72:04:55.02 & 2456491.9380 & 20.28(9)     & 10.26  & 0.05 & 145 & SR \\
EM2  & 00:24:06.269 & -72:04:45.36 & 2456530.7986 & 31.50(22)    &  9.98  & 0.08 & 140 & SR \\
EM3  & 00:24:08.310 & -72:04:50.69 & 2456847.9667 & 33.41(24)    & 10.07  & 0.05 & 134 & SR \\
EM4  & 00:24:07.203 & -72:04:46.45 & 2456498.9086 & 68.07(102)   & 13.89  & 0.14 & 149 & SR \\
EM5  & 00:24:05.087 & -72:04:54.38 &      --      &    --        & 10.34  & 0.06 & 145 & L \\
EM6  & 00:24:02.733 & -72:05:02.80 &      --      &    --        & 10.60  & 0.07 & 147 & L \\
EM7  & 00:24:07.885 & -72:05:02.00 &      --      &    --        &   --   & -- & 146 & NC \\

\hline
\end{tabular}
\end{table*}

\begin{figure}[ht]
\centering
\includegraphics[scale=1.0]{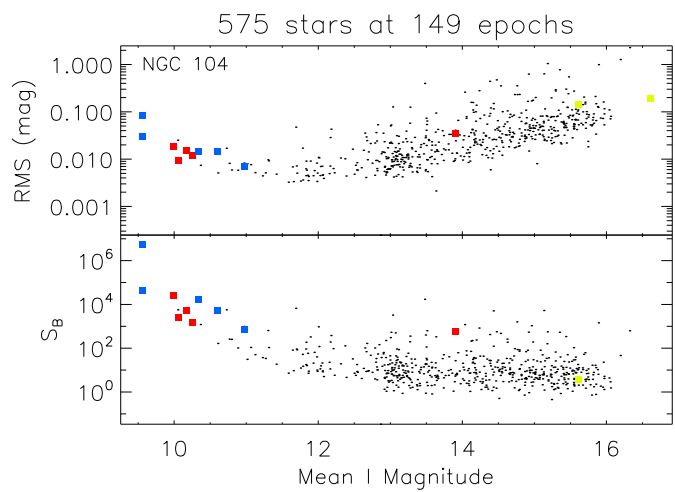}
\caption{Root mean square (RMS) magnitude deviation (top) and $S_B$ statistic (bottom) versus the 
mean $I$ magnitude for the 575 stars detected in the field of view of the reference image for 
NGC~104. Coloured points follow the convention adopted in Tab. \ref{tab:var_type} to identify 
the types of variables found in the field of this globular cluster.}
\label{fig:rms_sb_NGC0104}
\end{figure}

In Fig. \ref{fig:rms_sb_NGC0104}, the root mean square magnitude deviation (top) and S$_B$ statistic (bottom) are plotted versus the mean $I$ magnitude for a total of 575 light curves 
extracted in this analysis. The marked coloured points correspond to the variable stars studied in this work in comparison with the stars where no variation is found (normal black points).

\subsubsection{Known variables}

This globular cluster has of the order of 300 known variable sources in the Catalogue of Variable Stars in Galactic Globular Clusters 
\citep[version of summer 2007; ][]{clement01} which include long-period irregular variables, SX Phoenicis, RR Lyrae and binary systems. Most were found by 
\cite{albrow01+05}, \cite{weldrake04+03} and \cite{lebzelter05+01}. There are also 20 millisecond pulsars listed in the 
literature for this cluster \citep{freire01+05}. However, no visual counterparts have been found at their positions.

In the field of view covered by the reference image (Fig. \ref{fig:finding_chart_NGC104}) there are 49 previously known 
variables. However, for this cluster we were only able to detect stars brighter than $I$=16.1 mag (see RMS in Fig. 
\ref{fig:rms_sb_NGC0104}). Due to this, 5 known variable stars brighter than this limit were detected. These stars are 
labelled in the literature as PC1-V12, WF2-V34, LW10, LW11 and LW12. Our light curves for these variables can be found in 
Fig. \ref{fig:lc_chart1_NGC104}.

\textbf{PC1-V12}: This star was discovered by \cite{albrow01+05} and classified as a Blue Straggler star (BSS) with an average magnitude of V=16.076 mag. No period was reported in this case and it was not possible to find a period for this star in the present work.

\textbf{WF2-V34}: The variability of this star was found by \cite{albrow01+05} and was classified as a semi-regular variable with a 
period of P=5.5 d. However, using this period it was not possible to produce a good phased light curve 
and it was not possible to find another period for this star. Based on this, we classified this star as long-period irregular variable.

\textbf{LW10-LW12}: These three stars were discovered by \cite{lebzelter05+01}. Their positions in the colour-magnitude 
diagram (Fig. \ref{fig:cmd_NGC104}) confirm their cluster membership. \cite{lebzelter05+01} classified them as long-period irregular variables and suggested periods for these stars to be LW10: 110 d or 221 d; LW11: 36.0 d, and LW12 : 61 d or 116 d. However, LW10 
and LW12 are clearly irregular variables (see our Fig. 7, and Fig. 1 from \cite{lebzelter05+01}). For LW11 we found a period of 19.37 d and we 
classify it as semi-regular.

\subsubsection{New variables}

After extracting and analysing all possible variable sources in the field of our images, a total of 7 new variables (EM1-EM7) were 
found of which 4 are semi-regular variables, 2 are long-period irregular variables and 1 is unclassified.

For this cluster, the nomenclature employed for most of the variable stars discovered in previous works does not correspond to the typical numbering system (e. g. V1, V2, V3, ..., Vn). For example, variables discovered by \cite{lebzelter05+01} and \cite{weldrake04+03} are numbered using their initials LW and W, respectively; \cite{albrow01+05} used a PC or WF nomenclature making 
reference to the instrument employed during the observations. Due to this, we did not find it practical to assign the typical numbering system to the new variable stars discovered in this work and we decided to use as reference the EMCCD camera used in the observations. That is, new variables are numbered as EM1-EM7.

\begin{figure}[ht]
\centering
\includegraphics[scale=1.0]{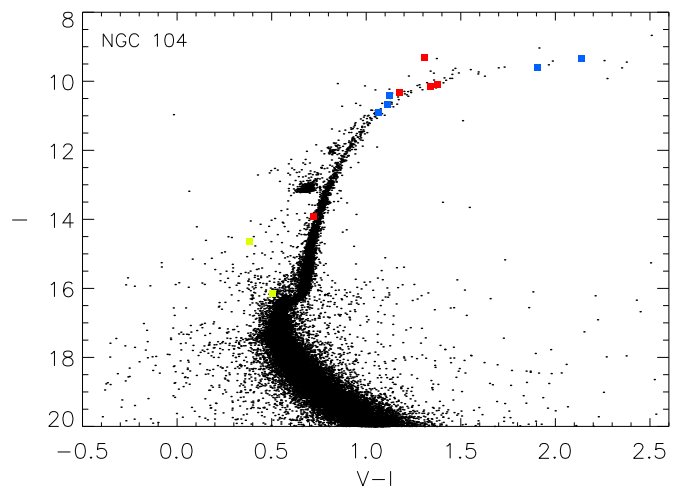}
\caption{Colour magnitude diagram for the globular cluster NGC~104 built with V and I magnitudes available in the ACS globular cluster survey extracted from HST images. The variable stars are plotted in colour following the convention 
adopted in Tab. \ref{tab:var_type}.}
\label{fig:cmd_NGC104}
\end{figure}

\begin{figure}[ht]
\centering
\includegraphics[scale=0.17]{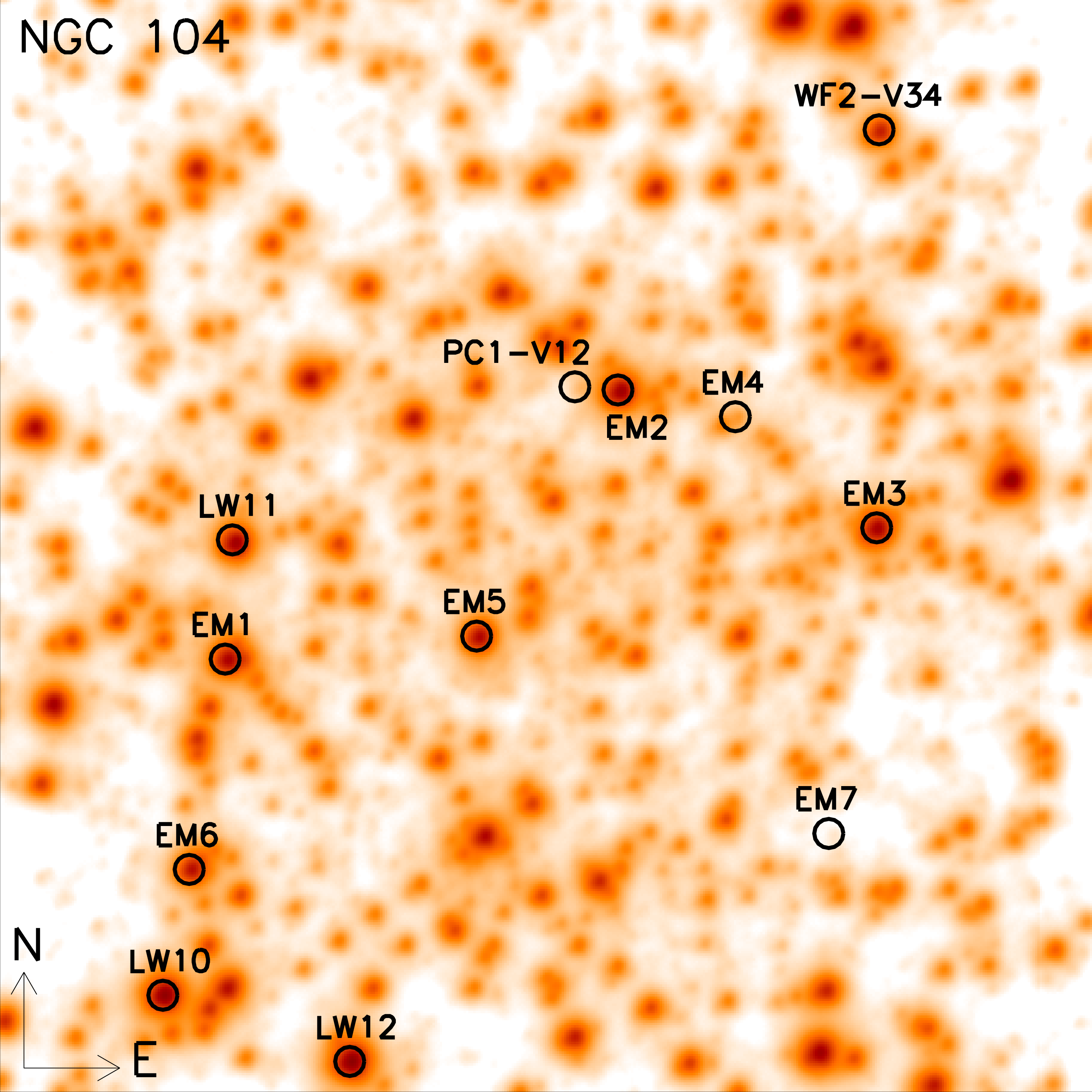}
\caption{Finding chart for the globular cluster NGC~104. The image used corresponds to the reference image constructed during the reduction. All known variables and new discoveries are labelled. Image size is $\sim41\times41$ arcsec$^2$.}
\label{fig:finding_chart_NGC104}
\end{figure}

\textbf{EM1-EM4}: These variable stars are placed at the top of the colour magnitude diagram (Fig. \ref{fig:cmd_NGC104}), their 
light curve shape, amplitudes and periods found suggest that these stars are semi-regular variables. Periods found for these 
variables are listed in Tab. \ref{tab:NGC104_ephemerides}. The star EM1 is not over the red giant branch in Fig. \ref{fig:cmd_NGC104} 
but instead it is more toward the asymptotic giant branch, it appears at the top on the left. Finally the star EM4 is located at the 
bottom part of the red giant branch just below the starting point of the horizontal branch.

\textbf{EM5-EM6}: These stars have amplitudes of $\sim$0.06 and 0.07 mag, respectively. They are located at the top of the red giant 
branch. We were not able to determine a proper period for these stars. Due to this, we classified them as long-period irregular variables.

\begin{figure*}[htp!]
\centering
\includegraphics[scale=1]{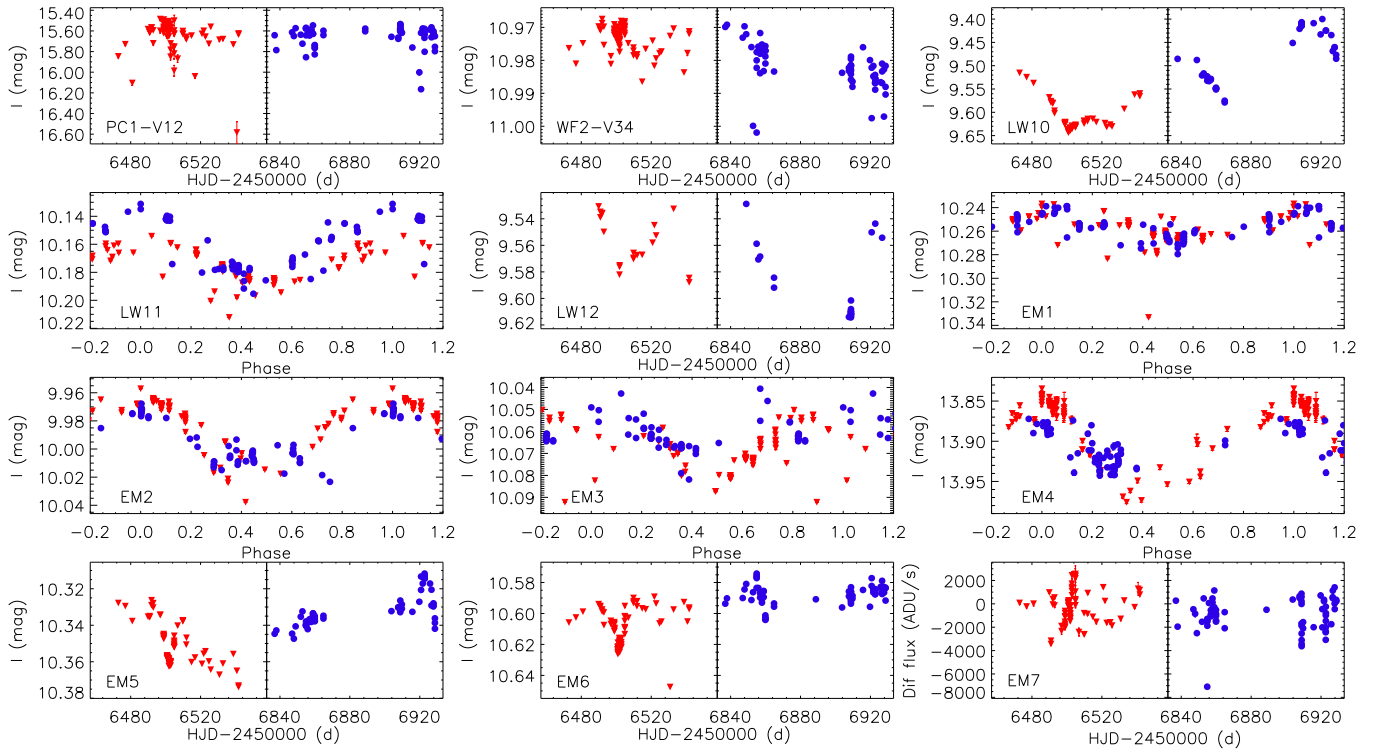}
  \caption{NGC~104: Light curves of the known and new variables discovered in this globular cluster. Red triangles correspond 
  to the data obtained during the year 2013 and blue circles correspond to the data obtained during the year 2014. For EM7, we plot the quantity $f_{\mathrm{diff}}(t)/p(t)$ since a reference flux 
is not available.}
\label{fig:lc_chart1_NGC104}
\end{figure*}

\textbf{EM7}: This star is placed in the Blue Straggler region. The variation in the difference images is clear for this star. However, 
the pipeline did not find this star in the reference image and no reference flux is available to convert the difference fluxes to magnitudes. 
The difference fluxes are plotted in Fig. \ref{fig:lc_chart1_NGC104}. We were unable to determine a period.

\subsection{\textbf{NGC~5139 / C1323-472/ Omega Centauri}}\label{sec:NGC5139}

This globular cluster was discovered by Edmond Halley in 1677\footnote{http://messier.seds.org/xtra/ngc/n5139.html}. The cluster 
is in the constellation of Centaurus at 5.2 kpc from the Sun and 6.4 kpc from the Galactic centre. It has a metallicity of 
[Fe/H]=-1.53 dex, a distance modulus of (m-M)$_V$=13.94 mag and the level of the horizontal branch is at V$_{HB}$=14.51 mag.

\begin{table*}[htp!]
\caption{NGC~5139: Ephemerides and main characteristics of the variable stars in the field of this globular cluster. Columns are the same as in Tab. \ref{tab:NGC104_ephemerides}.}
\label{tab:NGC5139_ephemerides}
\centering
\begin{tabular}{ccccccccc}
\hline\hline
Var id &   RA  &  Dec  &Epoch & $P$ & $I_{median}$ & $A_{\mathrm{i}^{\prime}+\mathrm{z}^{\prime}}$ & $N$ & Type\\
       & J2000 & J2000 & HJD  & d &      mag     &                mag                            &   &     \\
\hline
V457 & 13:26:46.246 & -47:28:44.81 & -- & -- & 15.82 & 0.15 & 78 & NC \\
V458 & 13:26:46.103 & -47:28:57.05 & -- & -- & 10.21 & 0.04 & 78 & L \\
V459 & 13:26:46.313 & -47:28:40.33 & -- & -- & 10.98 & 0.06 & 78 & L \\

\hline
\end{tabular}
\end{table*}

\begin{figure}[htp!]
\centering
\includegraphics[scale=1.0]{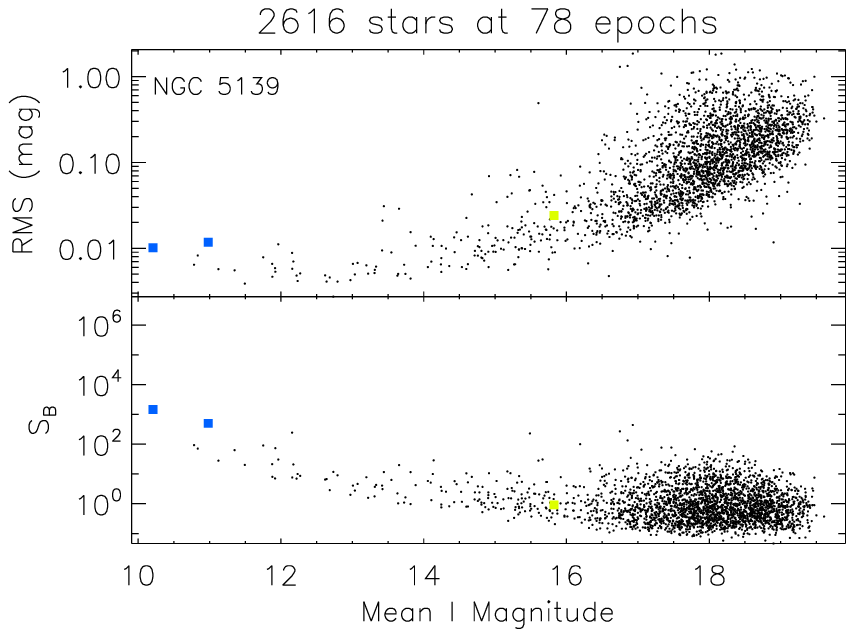}
\caption{Root mean square (RMS) magnitude deviation (top) and $S_B$ statistic (bottom) versus the 
mean $I$ magnitude for the 2616 stars detected in the field of view of the reference image for 
NGC~5139. Coloured points follow the convention adopted in Tab. \ref{tab:var_type} to identify 
the types of variables found in the field of this globular cluster.}
\label{fig:rms_sb_NGC5139}
\end{figure}

\subsubsection{Known variables}

In the Catalogue of Variable Stars in Galactic Globular Clusters \citep{clement01} there are of the order of 400 variable stars for this globular cluster. Most of these 
variables were discovered by \cite{bailey1902}, \cite{Kaluzny04+05} and \cite{weldrake07+02}. All of the known variables are 
outside the field of view of our reference image. Most recently, \cite{Navarrete15+09} made an updated analysis of the 
variables in this cluster but the variables in their study are also located outside the field of view of our reference image.

\subsubsection{New variables}

We have found 3 new variables in this globular cluster. Two are long-period irregular variables and one is unclassified.

\begin{figure}[ht]
\centering
\includegraphics[scale=1]{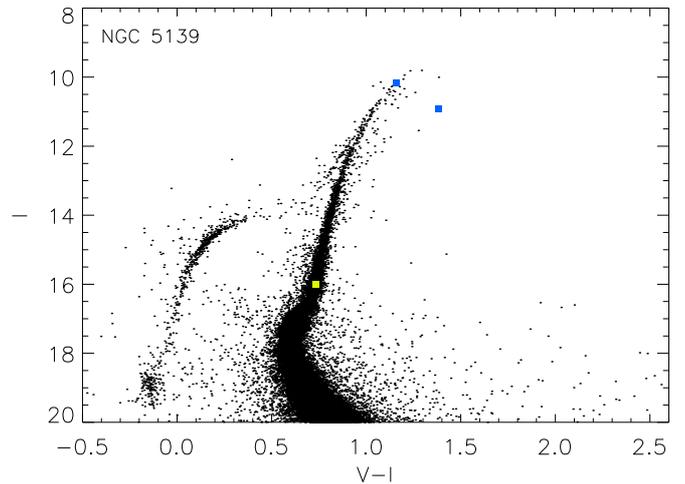}
\caption{Colour magnitude diagram of the globular cluster NGC~5139 built with V and I magnitudes available in the ACS globular cluster survey extracted from HST images. The variable stars are plotted in colour following the convention adopted in Tab. \ref{tab:var_type}.}
\label{fig:cmd_NGC5139}
\end{figure}

The new variable stars are plotted in the RMS diagram and $S_B$ statistic (see Fig. \ref{fig:rms_sb_NGC5139}). Based on the position 
of the stars in Figs. \ref{fig:rms_sb_NGC5139} and \ref{fig:cmd_NGC5139} there is no evidence of RR Lyrae in the field 
covered by this work neither in the inspection of the difference images obtained in the reductions.

\textbf{V457}: This variable star has an amplitude of $\sim$0.15 mag with a median magnitude $I=$15.82 mag. It lies on the RGB and we were unable to find a period.

\textbf{V458, V459}: Theses two stars are at the top of the red giant branch in Fig. \ref{fig:cmd_NGC5139} and have amplitudes of $\sim$0.04 and $\sim$0.06 mag, respectively. No periods were found in this work. V459 is on the red side of the main red giant branch. We classify them as long-period irregular variables.

\begin{figure}[ht]
\centering
\includegraphics[scale=0.17]{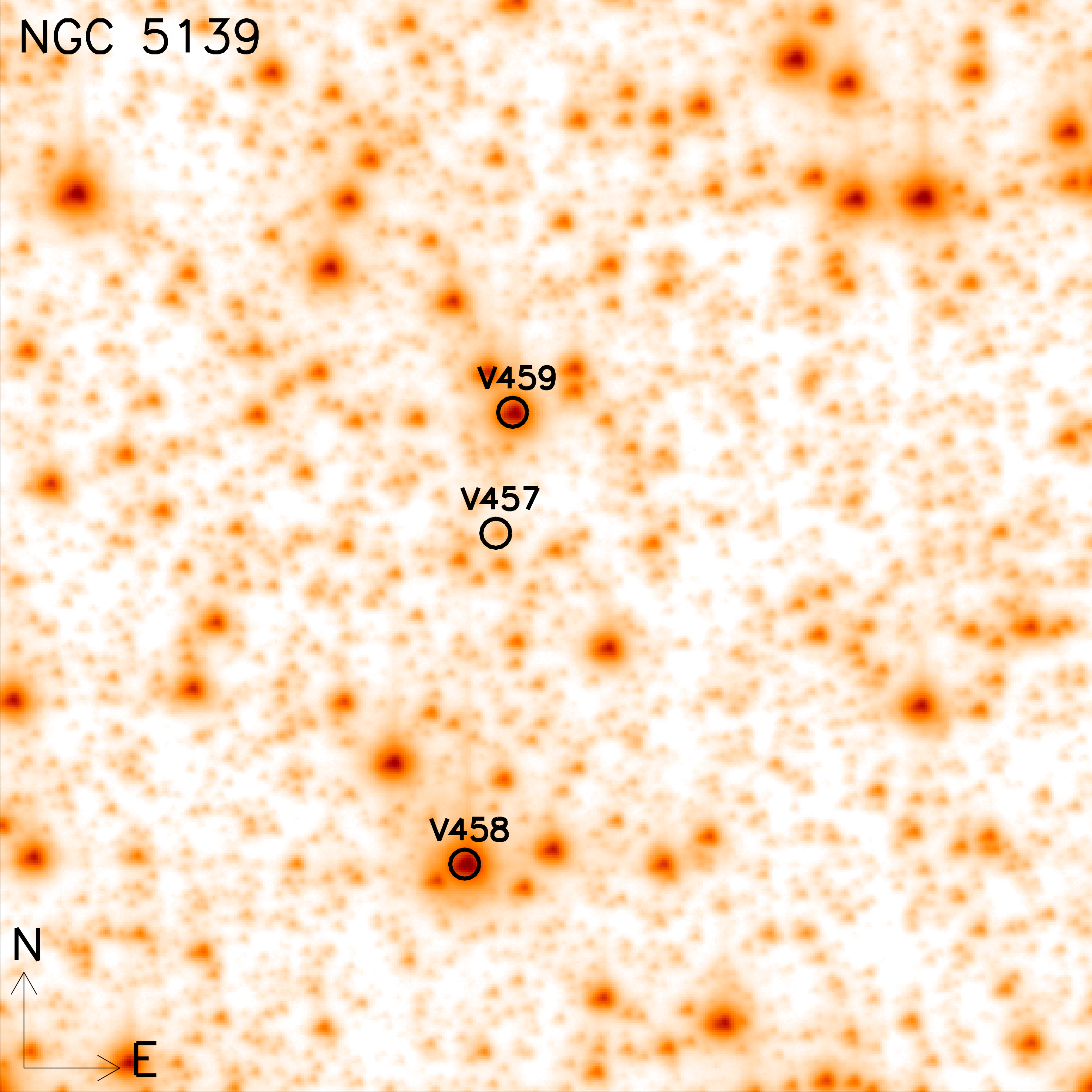}
\caption{Finding chart for the globular cluster NGC~5139. The image used corresponds to the reference image constructed during the reduction. The new variables discovered are labelled. Image size is $\sim41\times41$ arcsec$^2$.}
\label{fig:finding_chart_NGC5139}
\end{figure}

\begin{figure*}[htp!]
\centering
\includegraphics[scale=1 ]{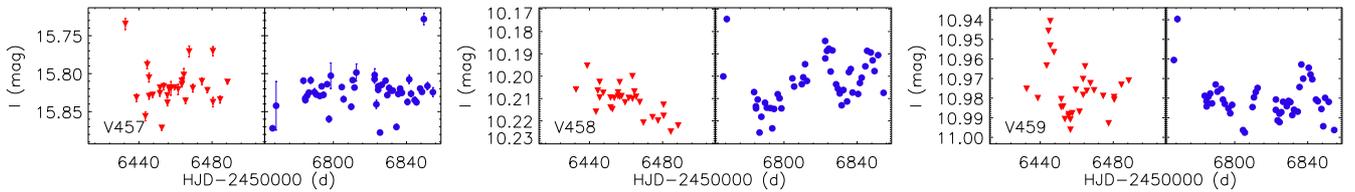}
  \caption{NGC~5139: Light curves of the 3 new variables discovered in this globular cluster. Symbols are the same as in Fig. \ref{fig:lc_chart1_NGC104}.}
\label{fig:lc_chart1_NGC5139}
\end{figure*}

\subsection{\textbf{NGC~5286 / C1343-511 / Caldwell 84}}\label{sec:NGC5286}

This globular cluster was discovered by James Dunlop in 1827 \citep{caldwellcatag}. It is in the constellation of Centaurus 
at 11.7 kpc from the Sun and 8.9 kpc from the Galactic centre. It has a metallicity of [Fe/H]=-1.69 dex, a distance 
modulus of (m-M)$_V$=16.08 mag and the magnitude level of the horizontal branch is at V$_{HB}$=16.63 mag.

\begin{figure}[ht]
\centering
\includegraphics[scale=1.0]{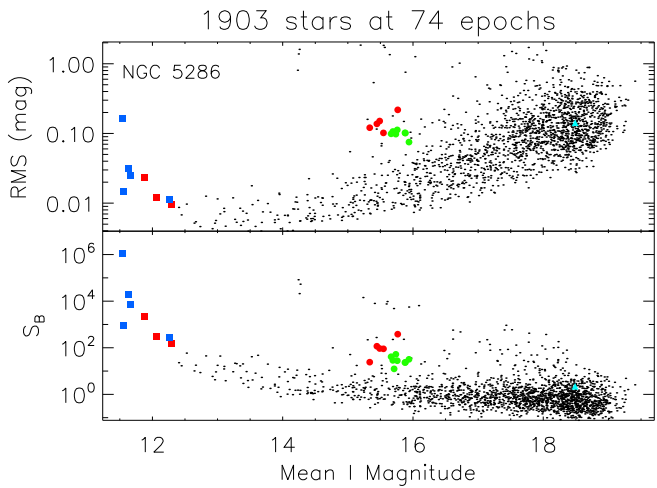}
\caption{Root mean square (RMS) magnitude deviation (top) and $S_B$ statistic (bottom) versus the 
mean $I$ magnitude for the 1903 stars detected in the field of view of the reference image for 
NGC~5286. Coloured points follow the convention adopted in Tab. \ref{tab:var_type} to identify 
the types of variables found in the field of this globular cluster.}
\label{fig:rms_sb_NGC5286}
\end{figure}

In Fig. \ref{fig:rms_sb_NGC5286}, the RMS diagram and the $S_B$ statistic are shown for the sample of 1903 stars analysed in this 
globular cluster. Most of them have 74 epochs. All variable stars studied in this work are plotted using the 
colour classification given in Tab. \ref{tab:var_type}.

\begin{table*}[htp!]
\caption{NGC~5286: Ephemerides and main characteristics of the variable stars in the field of this globular cluster. Columns are the same as in Tab. \ref{tab:NGC104_ephemerides}.}
\label{tab:NGC5286_ephemerides}
\centering
\begin{tabular}{ccccccccc}
\hline\hline
Var id &   RA  &  Dec  &Epoch & $P$ & $I_{median}$ & $A_{\mathrm{i}^{\prime}+\mathrm{z}^{\prime}}$ & $N$ & Type\\
       & J2000 & J2000 & HJD  & d &      mag     &                mag                            &   &     \\
\hline
V37 & 13:46:27.275 & -51:22:51.63 & 2456847.4835 & 0.583886(85)  & 15.92 & 0.81 & 52 & RR0 \\
V39 & 13:46:25.458 & -51:22:46.81 & 2456768.6823 & 0.742099(136) & 15.57 & 0.36 & 74 & RR0 \\
V40 & 13:46:26.917 & -51:22:44.21 & 2456455.5615 & 0.365961(33)  & 15.76 & 0.30 & 74 & RR1 \\
V41 & 13:46:26.577 & -51:22:42.92 & 2456822.5630 & 0.322262(26)  & 15.88 & 0.42 & 74 & RR1 \\
V43 & 13:46:26.730 & -51:22:38.08 & 2456447.5363 & 0.658478(107) & 15.47 & 0.50 & 74 & RR0 \\
V46 & 13:46:27.470 & -51:22:33.93 & 2456831.5372 & 0.682690(115) & 15.52 & 0.56 & 74 & RR0 \\
V50 & 13:46:25.984 & -51:22:30.32 & 2456457.5131 & 0.365145(33)  & 15.71 & 0.41 & 74 & RR1 \\
V55 & 13:46:25.403 & -51:22:25.79 & 2456841.5096 & 0.288925(21)  & 15.91 & 0.30 & 74 & RR1 \\
V56 & 13:46:25.112 & -51:22:24.34 & 2456783.6351 & 0.283202(20)  & 15.97 & 0.22 & 74 & RR1 \\
V57 & 13:46:27.785 & -51:22:20.69 & 2456833.6352 & 0.294964(22)  & 15.75 & 0.40 & 74 & RR1 \\
V58 & 13:46:26.627 & -51:22:19.29 & 2456463.5240 & 0.367064(33)  & 15.65 & 0.30 & 74 & RR1 \\
\hline
V63 & 13:46:25.405 & -51:22:44.92 & 2456811.5275 &  0.0486320(10)  & 18.52 & 0.55 & 74 & SXPhe \\
V64 & 13:46:26.780 & -51:22:26.60 & 2456833.6900 &  0.369783(34) & 15.72 & 0.41 & 74 & RR1 \\
V65 & 13:46:26.763 & -51:22:28.93 & 2456461.4946 &  0.619491(95) & 15.35 & \textbf{0.36} & 74 & RR0 \\
V66 & 13:46:27.166 & -51:22:40.14 & 2456833.7152 & 17.55(08)     & 12.06 & 0.05 & 74 & SR \\
V67 & 13:46:26.121 & -51:22:23.76 & 2456822.5630 & 21.20(11)     & 12.29 & 0.04 & 74 & SR \\
V68 & 13:46:26.190 & -51:22:26.55 & 2456844.4857 & 32.95(27)     & 11.87 & 0.10 & 74 & SR \\
V69 & 13:46:25.869 & -51:22:16.18 &      --      &     --        & 12.26 & 0.05 & 74 & L \\
V70 & 13:46:24.334 & -51:22:47.14 &      --      &     --        & 11.57 & 0.07 & 63 & L \\
V71 & 13:46:27.668 & -51:22:35.25 &      --      &     --        & 11.66 & 0.09 & 74 & L \\
V72 & 13:46:27.364 & -51:22:16.22 &      --      &     --        & 11.64 & 0.14 & 74 & L \\
V73 & 13:46:27.008 & -51:22:29.88 &      --      &     --        & 11.51 & 0.52 & 74 & L \\

\hline
\end{tabular}
\end{table*}

\subsubsection{Known variables}

This globular cluster has 58 known variable stars listed in the Catalogue of Variable Stars in Galactic Globular Clusters \citep{clement01} of which 52 
are RR Lyrae stars. There are only 11 previously known variable stars in the field of view of our reference image (V37, V39, V40, V41, V43, V46, V50, V55, V56, V57, V58). All of them are RR Lyrae discovered by \cite{Zorotovic10+20} using DIA on imaging data from a one-week observing run.

The celestial coordinates given in Tab. 1 of \cite{Zorotovic10+20} do not match the positions of the known variable stars in the 
field of our images. We therefore used the finding chart given in their Fig. 1 to do a visual matching of the variables. As pointed out in the Catalogue of Variable Stars in Galactic Globular Clusters \citep{clement01}, there is a difference in the position of the variables studied by \cite{Zorotovic10+20} with respect to the position of the variables in \cite{Samus09+04} which is $\sim$6 arcseconds in declination and $\lesssim$1 arcsecond in right ascension. This difference is corroborated by the position of the variables in our reference image. Celestial coordinates of the positions we used are given in Tab. \ref{tab:NGC5286_ephemerides}.

Our extended observational baseline has allowed us to greatly improve the periods of the variables discovered by \cite{Zorotovic10+20}. Our 
period estimates are listed in Tab. \ref{tab:NGC5286_ephemerides} and have typical errors of 0.00002 - 0.00010 d. We confirm the variable star classifications made by \cite{Zorotovic10+20}.

Note that V41 is a strong blend with a brighter star that is only just resolved in our high-resolution reference image.

\subsubsection{New variables}

In this globular cluster we found 11 previously unknown variables where 5 are long-period irregular variables, 3 are semi-regular variables, 2 are RR Lyrae, and 1 is a SX Phoenicis.

\textbf{V63}: The star has a median magnitude of $I\sim$18.52 mag and it is in the blue straggler region. Its amplitude is 
$A_{\mathrm{i}^{\prime}+\mathrm{z}^{\prime}}\sim$0.49 mag and it has a period of P=0.0486320 d. The star is clearly a SX Phoenicis. The variation of this star was not 
detected using the RMS or $S_B$ statistic, although it is very clear in the difference images.

\textbf{V64}: This star is an RR Lyrae pulsating in the first overtone 
(RR1) with a period of 0.369783 d and an amplitude of 0.41 mag. In Fig. \ref{fig:finding_chart_NGC5286}, notice that V64 is \textbf{very close} to a bright star (6.808 pixels or 0.613 arcsec). This could be the reason why this variable was 
not discovered before and makes a good example of the benefits of using the EMCCD cameras and the shift-and-add technique along with DIA.

\textbf{V65}: This star is another RR Lyrae which is pulsating in the fundamental mode (RR0) with a period of 0.619491 d 
and an amplitude of $\sim$0.36 mag.

\begin{figure}[ht]
\centering
\includegraphics[scale=1]{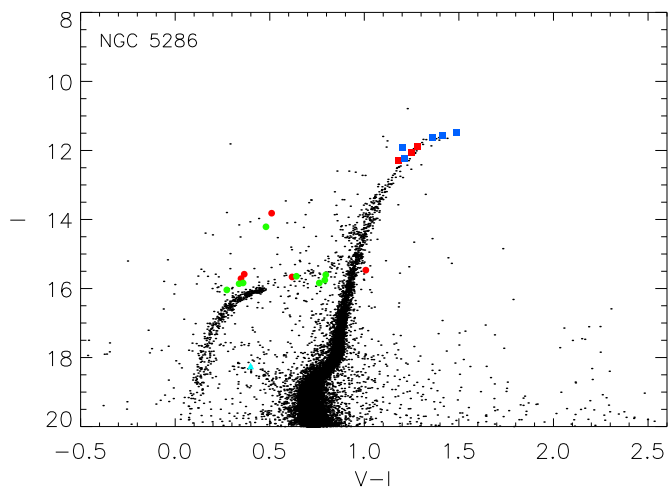}
\caption{Colour magnitude diagram for the globular cluster NGC~5286 built with V and I magnitudes available in the ACS globular cluster survey extracted from HST images. The variable stars are plotted in colour following the convention 
adopted in Tab. \ref{tab:var_type}.}
\label{fig:cmd_NGC5286}
\end{figure}

\begin{figure}[ht]
\centering
\includegraphics[scale=0.22]{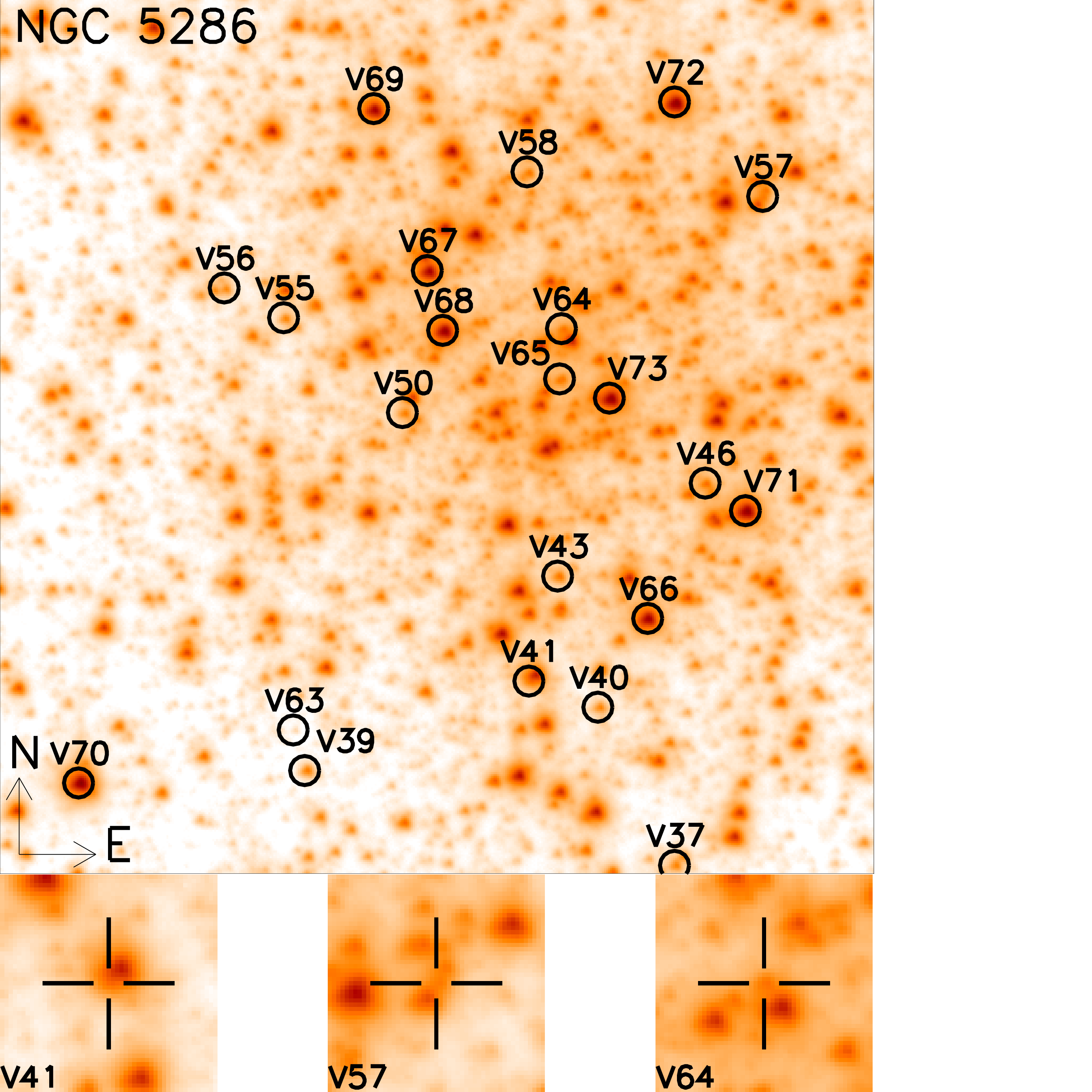}
\caption{Finding chart for the globular cluster NGC~5286. The image used corresponds to the reference image constructed during the reduction. All known variables and new discoveries are labelled. Image size is $\sim41\times41$ arcsec$^2$. The image stamps are of size $\sim4.6\times4.6$ arcsec$^2$.}
\label{fig:finding_chart_NGC5286}
\end{figure}

\textbf{V66-V68}: These stars are semi-regular variables. As it is seen in Fig. \ref{fig:cmd_NGC5286}, they are at the top 
of the red giant branch. They have amplitudes that range between 0.04 to 0.10 mag. They have periods between 
$\sim$17 to 33 d. Ephemerides for these stars can be found in Tab. \ref{tab:NGC5286_ephemerides}.

\textbf{V69-V73}: These five stars are also positioned at the top of the red giant branch with amplitudes of 0.05 to 0.52 mag. It 
was not possible to find periods for these stars in this work. Due to this, they were classified as long-period irregular variables.

In Fig. \ref{fig:apd_NGC5286} the amplitude-period diagram for the RR Lyrae stars studied in this cluster is shown. The filled lines 
correspond to the Oosterhoff type I (OoI) and the dashed lines correspond to the Oosterhoff type II (OoII) models defined by 
\cite{kunder13+04}. All RR0 variables (with exception of V65) fall on the model for OoII type while the RR1 stars scatter around both models. In this 
diagram the RR0 stars suggest an OoII type classification for NGC~5286 which is in agreement with the study done by \cite{Zorotovic10+20} where 
they found that their research pointed to an OoII status as well.

\begin{figure*}[htp!]
\centering
\includegraphics[scale=1 ]{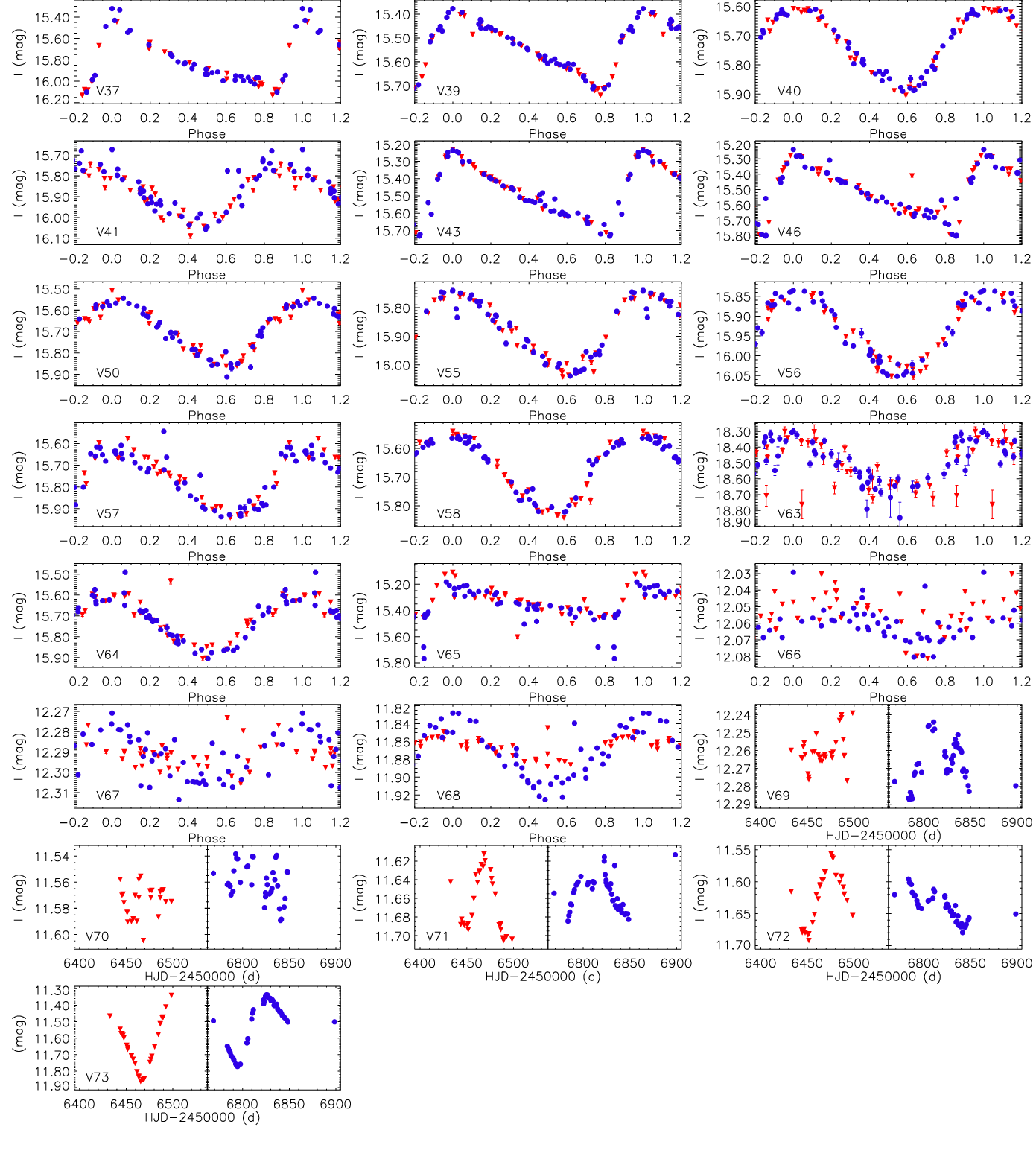}
  \caption{NGC~5286: Light curves of the known and new variables discovered in this globular cluster. Symbols are the same as in Fig. \ref{fig:lc_chart1_NGC104}.}
\label{fig:lc_chart1_NGC5286}
\end{figure*}

\begin{figure}[htp!]
\centering
\includegraphics[scale=1 ]{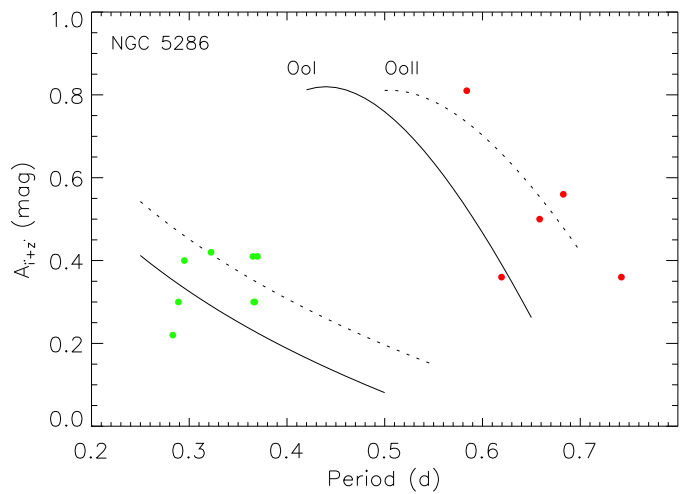}
\caption{Amplitude-period diagram for the RR Lyrae stars in the globular cluster NGC~5286.}
\label{fig:apd_NGC5286}
\end{figure}

\subsection{\textbf{NGC~6093 / C1614-228 / M 80}}\label{sec:NGC6093}

This globular cluster was discovered by Charles Messier in 1781\footnote{http://messier.seds.org/m/m080.html}. This 
cluster is in the constellation of Scorpius at 10.0 kpc from the Sun and 3.8 kpc from the Galactic centre. Its has a 
metallicity of [Fe/H]=-1.75 dex, a distance modulus of $\mathrm{(m-M)}_V$=15.56 mag and the position of its horizontal branch 
is at V=16.10 mag.

\begin{table*}[htp!]
\caption{NGC~6093: Ephemerides and main characteristics of the variable stars in the field of this globular cluster. Columns are the same as in Tab. \ref{tab:NGC104_ephemerides}.}
\label{tab:NGC6093_ephemerides}
\centering
\begin{tabular}{ccccccccc}
\hline\hline
Var id &   RA  &  Dec  &Epoch & $P$ & $I_{median}$ & $A_{\mathrm{i}^{\prime}+\mathrm{z}^{\prime}}$ & $N$ & Type\\
       & J2000 & J2000 & HJD  & d &      mag     &                mag                            &   &     \\
\hline
V10 & 16:17:01.163 & -22:58:34.57 & 2456805.5870 & 0.614154(92)  & 14.25 & 0.29 & 52 & RR0 \\
V17 & 16:17:02.864 & -22:58:32.55 & 2456831.5535 & 0.415081(42)  & 15.31 & 0.31 & 62 & RR1 \\
V19 & 16:17:02.107 & -22:58:29.10 & 2456844.5144 & 0.596064(87)  & 15.28 & 0.54 & 62 & RR0 \\
V20 & 16:17:03.263 & -22:58:37.49 & 2456831.5535 & 0.745207(136) & 15.31 & 0.46 & 62 & RR0 \\
V34 & 16:17:02.820 & -22:58:33.80 &      --      &      --       & 16.44 & 0.36 & 62 & CV(Nova) \\
\hline
V35 & 16:17:03.313 & -22:58:33.15 &      --      &      --       & 11.40 & 0.09 & 62 & L \\
V36 & 16:17:03.145 & -22:58:41.92 &      --      &      --       & 11.50 & 0.12 & 62 & L \\
V37 & 16:17:02.320 & -22:58:30.48 &      --      &      --       & 11.51 & 0.09 & 62 & L \\
V38 & 16:17:03.263 & -22:58:34.96 &      --      &      --       & 11.70 & 0.04 & 62 & L \\
V39 & 16:17:02.201 & -22:58:34.49 &      --      &      --       & 11.75 & 0.07 & 62 & L \\
V40 & 16:17:03.042 & -22:58:25.72 &      --      &      --       & 12.36 & 0.13 & 62 & L \\

\hline
\end{tabular}
\end{table*}

\begin{figure}[ht]
\centering
\includegraphics[scale=1.0]{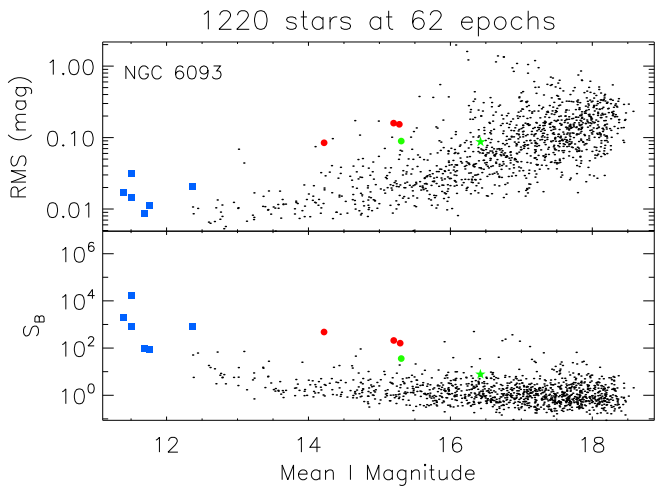}
\caption{Root mean square (RMS) magnitude deviation (top) and $S_B$ statistic (bottom) versus the 
mean $I$ magnitude for the 1220 stars detected in the field of view of the reference image for 
NGC~6093. Coloured points follow the convention adopted in Tab. \ref{tab:var_type} to identify 
the types of variables found in the field of this globular cluster.}
\label{fig:rms_sb_NGC6093}
\end{figure}

A total of 1220 light curves were extracted in the field covered by the reference image for the globular cluster NGC~6093. 
Most of the stars have 58 epochs in their light curves and variable stars studied in this work are plotted with colour in 
the RMS and $S_B$ diagrams shown in Fig. \ref{fig:rms_sb_NGC6093}.

\subsubsection{Known variables}

This globular cluster has 34 known variable sources in the Catalogue of Variable Stars in Galactic 
Globular Clusters \citep{clement01}. Only 7 are in the field of view covered by these observations. 
There is no visual detection of V11 in our images \cite[discovered by][and classified as a possible 
U Geminorum-type cataclysmic variable]{shara05+02}. This is to be expected since it has a U mean 
magnitude of 19.5 mag. The star V33 was discovered by \cite{dieball10+04} in an ultraviolet survey 
using the Hubble Space Telescope. They reported a UV magnitude for this target of about 19.8 mag. 
Again it was not possible to obtain a light curve for this faint target.

For the remaining variable stars in the field of view, three of them are RR0 (V10, V19 and V20), one 
is a RR1 (V17), and V34 is a Nova. Light curves for these stars are presented in Fig. 
\ref{fig:lc_chart1_NGC6093}. We have improved on the period estimates published by 
\cite{kopacki13} for the four RR Lyrae variables since we observed them using a time baseline of over 1 year, 
whereas \cite{kopacki13} analysed observations spanning only one week.

\textbf{V34:} This is the Nova discovered by Arthur von Auwers at Koenigsberg Observatory on May 21, 1860 \citep{luther1860}. As pointed out in the Catalogue of Variable Stars in Galactic Globular Clusters 
\citep{clement01}, an account of its discovery was given by \cite{sawyer38} in which a maximum visual apparent magnitude of 6.8~mag was reported using the data taken by von Auwers and Luther. Another review can also be found in \cite{wehlau90}. However, no observations of the Nova in outburst have been made until the present study. During our 2013 observing campaign, we caught an outburst of amplitude $\sim$0.36 mag starting at around HJD$\sim$2456500 d which lasted for $\sim$50 days. In our 2014 data the Nova continued at its baseline magnitude at $\sim$16.44 mag.

\cite{dieball10+04} in their ultraviolet survey assigned the label CX01 to an X-ray source that was found to be associated with the position of the Nova at the coordinates RA(J2000)=16:17:02.817 and DEC(J2000)=-22:58:33.92. These coordinates match with the position 
of the outburst found in this work and details are given in Tab. \ref{tab:NGC6093_ephemerides}.

\begin{figure}[ht]
\centering
\includegraphics[scale=1]{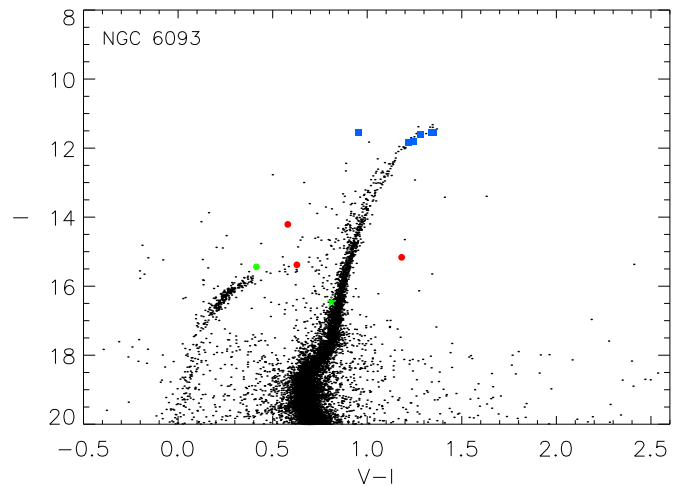}
\caption{Colour magnitude diagram for the globular cluster NGC~6093 built with V and I magnitudes available in the ACS globular cluster survey extracted from HST images. The variable stars are plotted in colour following the convention 
adopted in Tab. \ref{tab:var_type}.}
\label{fig:cmd_NGC6093}
\end{figure}

The position in the colour magnitude diagram shown in Fig. \ref{fig:cmd_NGC6093} suggests that this system is a cluster member. It is located at the bottom part of the red giant branch. Its position 
is plotted with a green five pointed star.

As the Nova with its outburst has shown variability, we have assigned the label V34 to this object.

\begin{figure}[ht]
\centering
\includegraphics[scale=0.17]{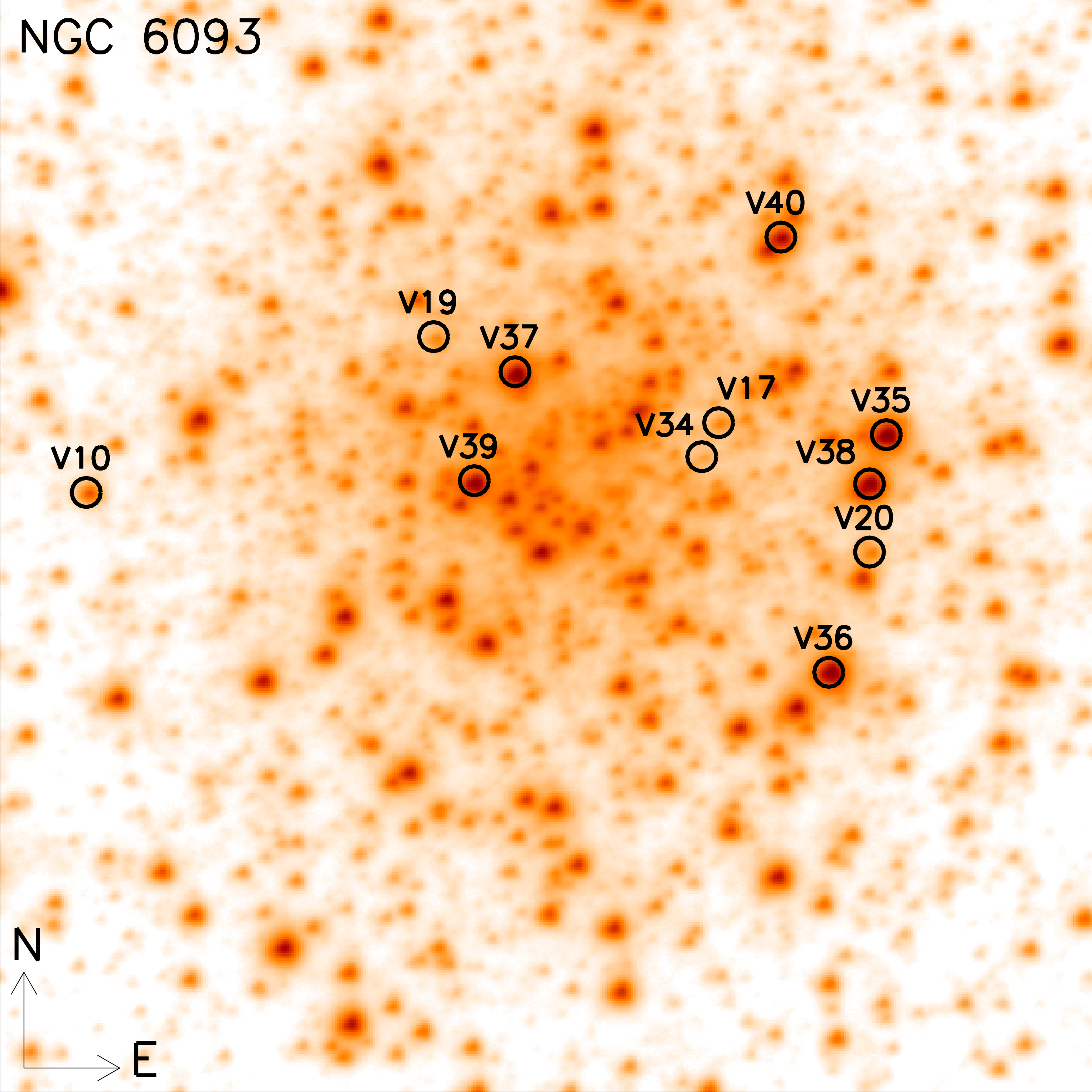}
\caption{Finding chart for the globular cluster NGC~6093. The image used corresponds to the reference image constructed during the reduction. All known variables and new discoveries are labelled. Image size is $\sim41\times41$ arcsec$^2$.}
\label{fig:finding_chart_NGC6093}
\end{figure}

\subsubsection{New variables}

A total of 6 new variables were found in this work. All of them are long-period irregular variables.

\textbf{V35-V40}: These stars are long-period irregular variables. They are located at the top of the red giant branch (see 
Fig. \ref{fig:cmd_NGC6093}). Their amplitudes go from $\sim$0.04 to 0.13 mag. The star V40 is the one 
placed at the blue side of the red giant branch. We did not find any kind of regular periodicity in the variation 
of these stars.

\begin{figure*}[htp!]
\centering
\includegraphics[scale=1 ]{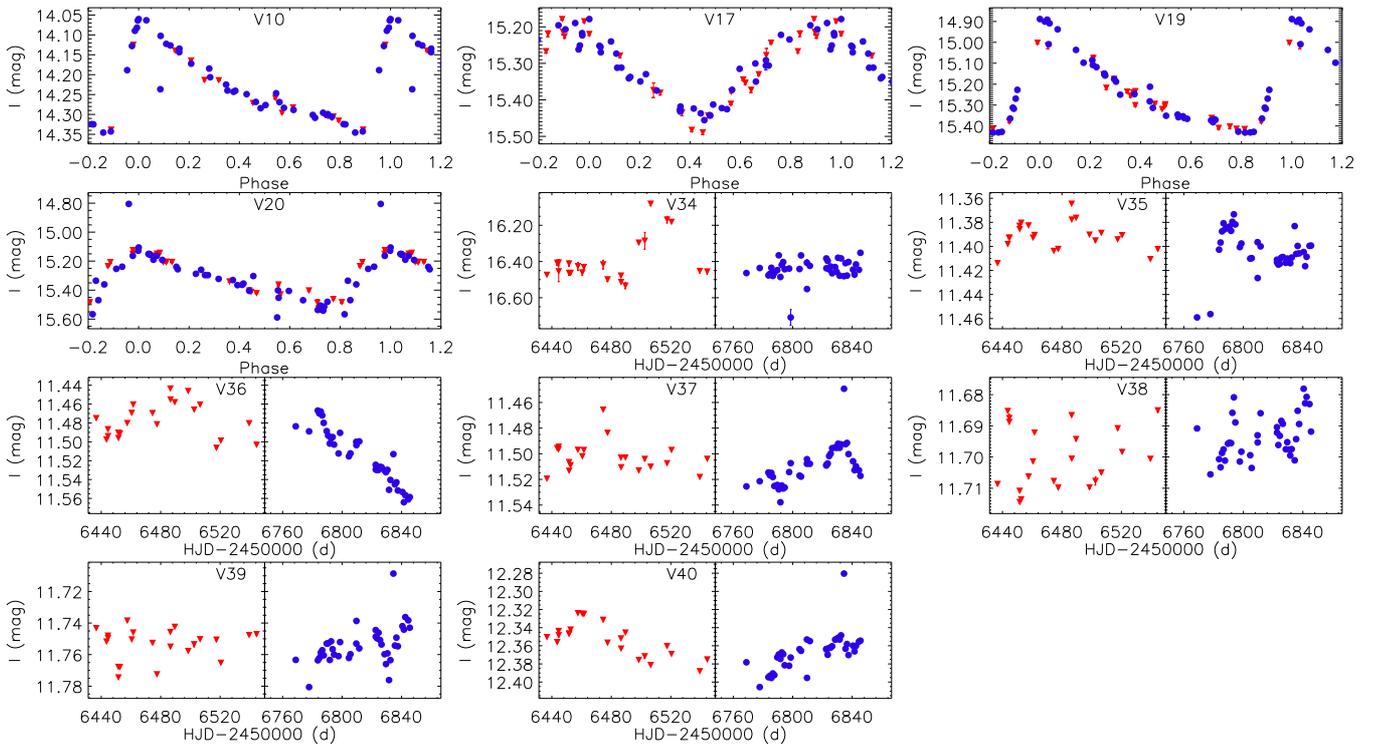}
  \caption{NGC~6093: Light curves of the known and new variables discovered in this globular cluster. Symbols are the same as in Fig. \ref{fig:lc_chart1_NGC104}.}
\label{fig:lc_chart1_NGC6093}
\end{figure*}

\subsection{\textbf{NGC~6121 / C1620-264 / M4}}\label{sec:NGC6121}

This globular cluster was discovered by Philippe Loys de Ch\'eseaux in 1746\footnote{http://messier.seds.org/m/m004.html}. This cluster is 
in the constellation of Scorpius at a distance of 2.2 kpc from the Sun and 5.9 kpc from the Galactic centre. It has a metallicity of [Fe/H]=-1.16 dex 
and a distance modulus of (m-M)$_V$=12.82 mag. Its horizontal branch level is at V$_{HB}$=13.45 mag.

\begin{table*}[htp!]
\caption{NGC~6121: Ephemerides and main characteristics of one variable star in the field of this globular cluster. Columns are the same as in Tab. \ref{tab:NGC104_ephemerides}.}
\label{tab:NGC6121_ephemerides}
\centering
\begin{tabular}{ccccccccc}
\hline\hline
Var id &   RA  &  Dec  &Epoch & $P$ & $I_{median}$ & $A_{\mathrm{i}^{\prime}+\mathrm{z}^{\prime}}$ & $N$ & Type\\
       & J2000 & J2000 & HJD  & d &      mag     &                mag                            &   &     \\
\hline
V21 & 16:23:35.903 & -26:31:33.68 & 2456477.5041 & 0.472008(56) & 12.55 & 0.92 & 64 & RR0 \\

\hline
\end{tabular}
\end{table*}

This globular cluster has of the order of 100 known variables in the Catalogue of Variable Stars in Galactic Globular Clusters \citep{clement01} 
and only three are in the field of view of the reference image; namely V21, V81 and V101. Furthermore, only V21 
is bright enough to be detected. The light curve for this RR0 star is shown in 
Fig. \ref{fig:lc_chart1_NGC6121}. The known period P=0.4720 d produces a very well phased light curve and it is in agreement with 
the period P=0.472008 d found in the analysis of this star.

\begin{figure}[ht]
\centering
\includegraphics[scale=0.17]{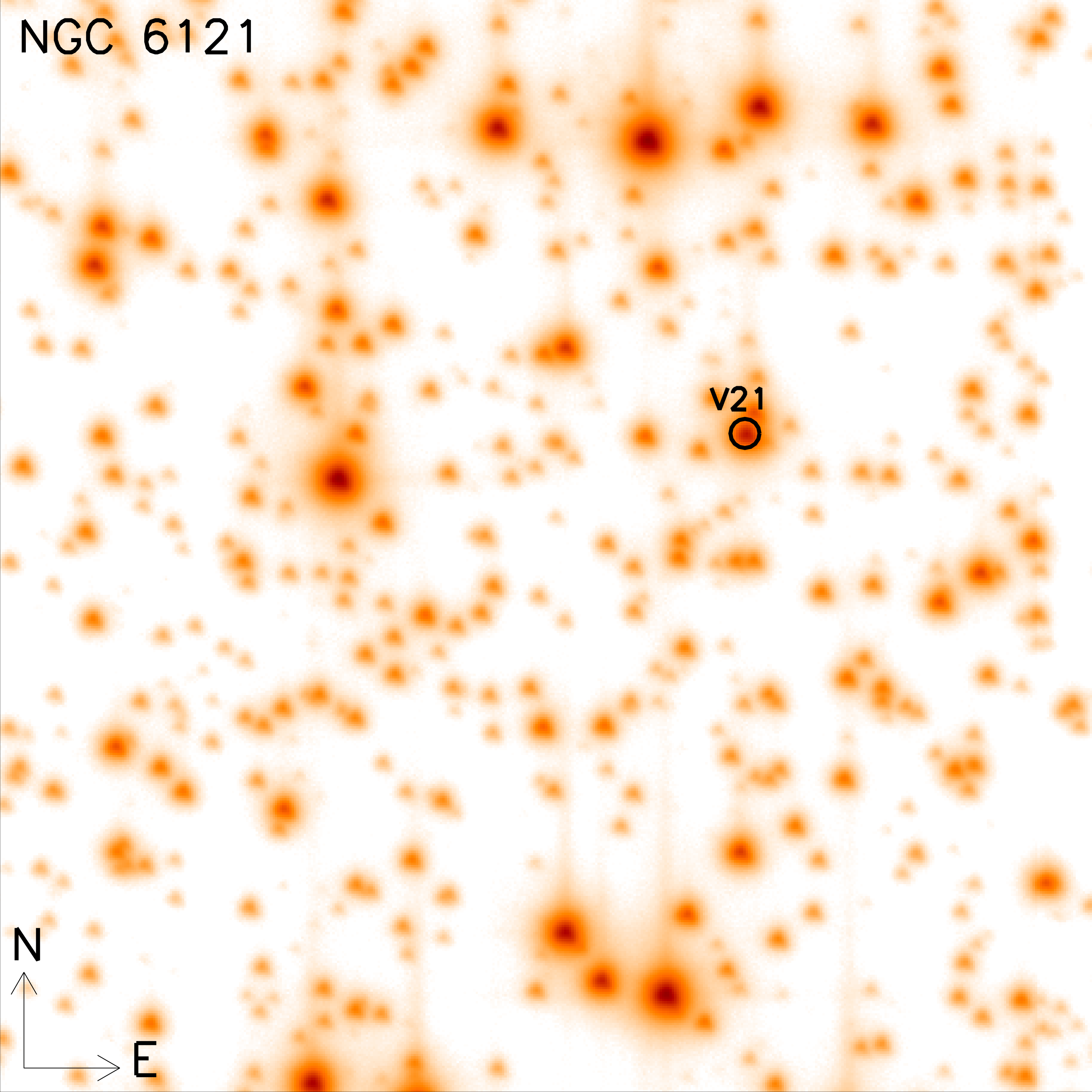}
\caption{Finding chart for the globular cluster NGC~6121. The image used corresponds to the reference image constructed during the reduction. The only known variable in the field is labelled. Image size is $\sim41\times41$ arcsec$^2$.}
\label{fig:finding_chart_NGC6121}
\end{figure}

No new variable stars were found in the field covered by the reference image for this globular cluster.

\begin{figure}[htp!]
\centering
\includegraphics[scale=1.07]{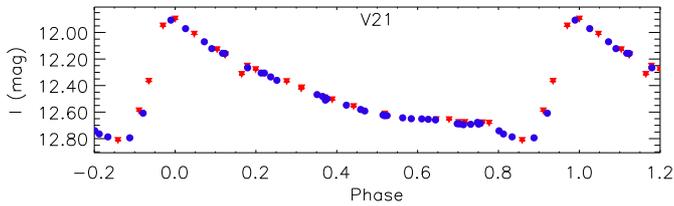}
  \caption{NGC~6121: Light curve of the variable V21 in this globular cluster. Symbols are the same as in Fig. \ref{fig:lc_chart1_NGC104}.}
\label{fig:lc_chart1_NGC6121}
\end{figure}

\subsection{\textbf{NGC~6541 / C1804-437}}\label{sec:NGC6541}

This globular cluster was discovered by N. Cacciatore in 1826\footnote{http://spider.seds.org/spider/MWGC/n6541.html}. It 
is in the constellation of Corona Australis at 7.5 kpc from the Sun and 2.1 kpc from the Galactic centre. The cluster has a 
metallicity of [Fe/H]=-1.81 dex, a distance modulus of $\mathrm{(m-M)}_V$=14.82 mag and the level of the horizontal branch is at V$_{HB}$=15.35 mag.

\begin{table*}[htp!]
\caption{NGC~6541: Ephemerides and main characteristics of the variable stars in the field of this globular cluster. Columns are the same as in Tab. \ref{tab:NGC104_ephemerides}.}
\label{tab:NGC6541_ephemerides}
\centering
\begin{tabular}{ccccccccc}
\hline\hline
Var id &   RA  &  Dec  &Epoch & $P$ & $I_{median}$ & $A_{\mathrm{i}^{\prime}+\mathrm{z}^{\prime}}$ & $N$ & Type\\
       & J2000 & J2000 & HJD  & d &      mag     &                mag                            &   &     \\
\hline
V14 & 18:08:01.670 & -43:42:52.26 & 2456908.7354 & 0.064669(1)  & 16.38 & 0.27 & 43 & SXPhe \\
\hline
V21 & 18:08:02.180 & -43:42:52.91 & 2456782.8812 & 0.596000(76) & 14.39 & 0.59 & 43 & RR0 \\
V22 & 18:08:03.743 & -43:42:34.46 &      --      &      --      & 10.74 & 0.17 & 29 & L \\

\hline
\end{tabular}
\end{table*}

\begin{figure}[ht]
\centering
\includegraphics[scale=1.0]{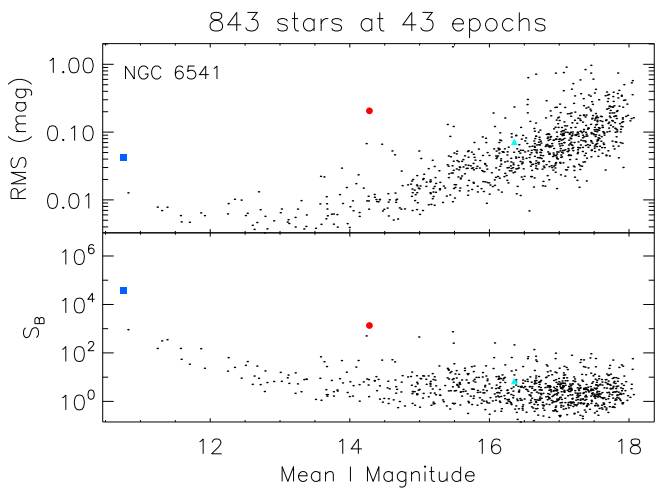}
\caption{Root mean square (RMS) magnitude deviation (top) and $S_B$ statistic (bottom) versus the 
mean $I$ magnitude for the 843 stars detected in the field of view of the reference image for 
NGC~6541. Coloured points follow the convention adopted in Tab. \ref{tab:var_type} to identify 
the types of variables found in the field of this globular cluster.}
\label{fig:rms_sb_NGC6541}
\end{figure}

A total of 843 light curves were extracted in this globular cluster. Most of them have 42 epochs. RMS diagram and $S_B$ statistic are shown in Fig. \ref{fig:rms_sb_NGC6541}.

\subsubsection{Known variables}

This globular cluster has 20 known variables. Four of them are in the field of view of our reference image: V12, V14, V15, and V17. \cite{fiorentino14+05} discovered and classified these stars as SX Phoenicis in their study carried out using observations with the HST. We were only able to recover V14 in our data around 2 arcseconds in declination from the reported position for this star in \cite{fiorentino14+05}. We did not detect variability at the positions given for V12, V15 and V17 neither in their surrounding areas, more likely due to the fact that the stars are too faint to be detected in our reference image. A reference image with higher signal to noise ratio might be needed in future studies of this globular cluster and longer exposure times for each observation.

\textbf{V14:} For this star we refined the period estimate to 0.064669 d from the previously listed value of 0.0649 d in \cite{fiorentino14+05}.

\begin{figure}[ht]
\centering
\includegraphics[scale=1]{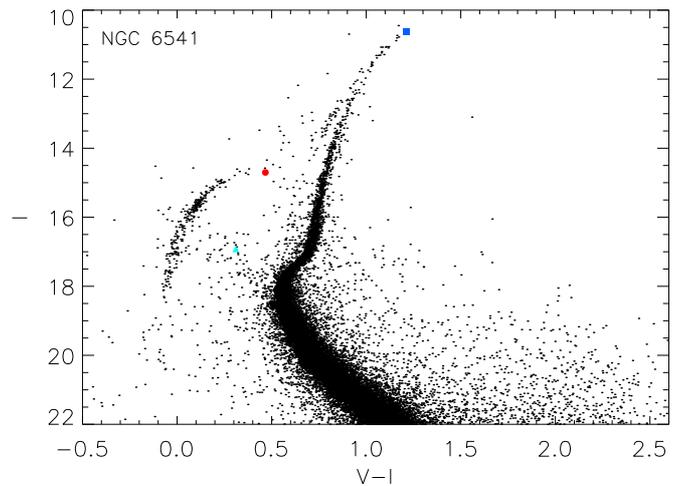}
\caption{Colour magnitude diagram for the globular cluster NGC~6541 built with V and I magnitudes available in the ACS globular cluster survey extracted from HST images. The variable stars are plotted in colour following the convention 
adopted in Tab. \ref{tab:var_type}.}
\label{fig:cmd_NGC6541}
\end{figure}

\begin{figure}[ht]
\centering
\includegraphics[scale=0.17]{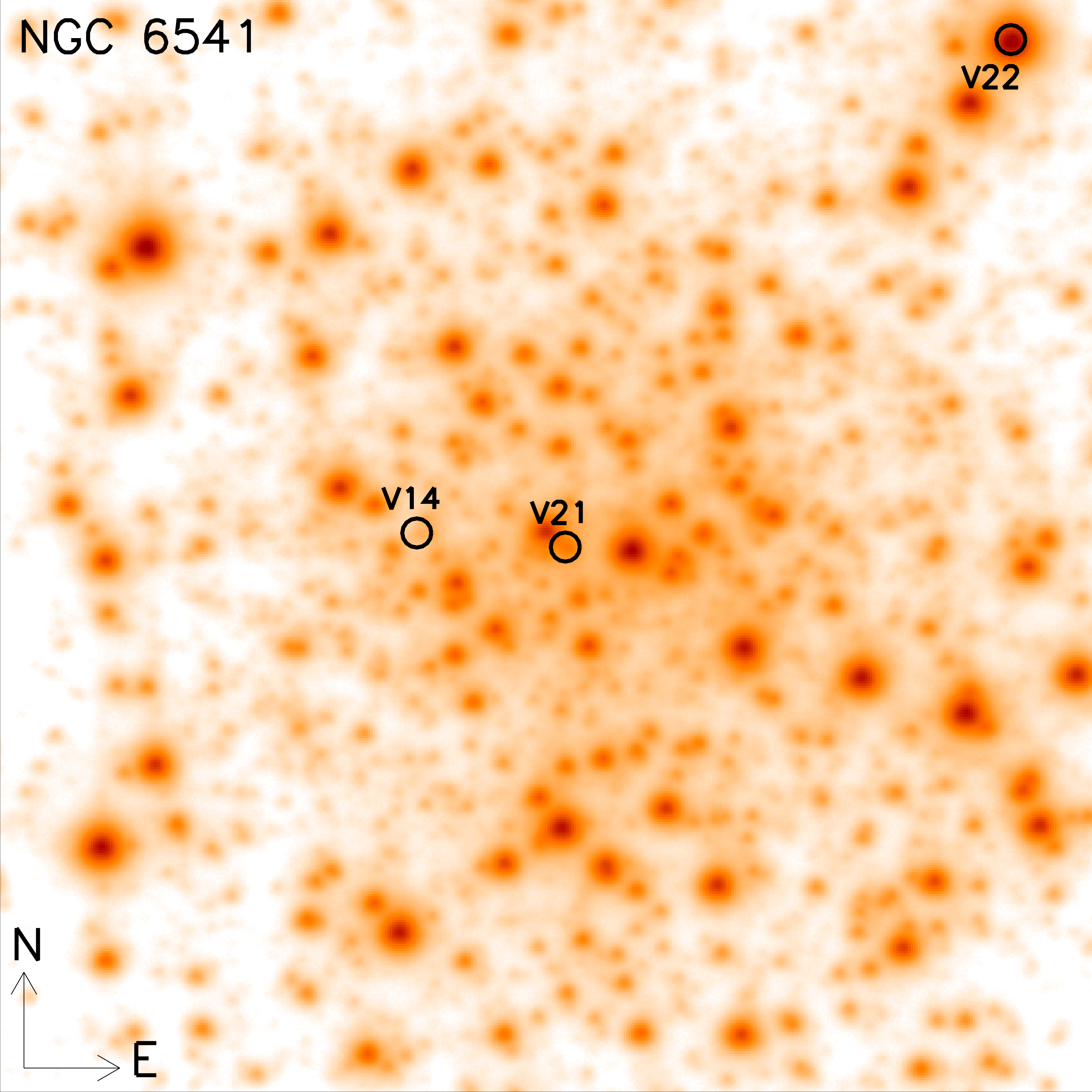}
\caption{Finding chart for the globular cluster NGC~6541. The image used corresponds to the reference image constructed during the reduction. All known variables and new discoveries are labelled. Image size is $\sim41\times41$ arcsec$^2$.}
\label{fig:finding_chart_NGC6541}
\end{figure}

\subsubsection{New variables}

A total of two new variable stars were found in this globular cluster: one RR Lyrae and one long-period irregular variable.

\textbf{V21:} This star is a RR Lyrae pulsating in the fundamental mode. In the colour-magnitude diagram (Fig. \ref{fig:cmd_NGC6541}) for this cluster it falls exactly in the instability strip of the horizontal branch. It is 
next to a brighter star ($\sim$1.039 arcsec) and this may be why it was not discovered before.

\textbf{V22:} This star is at the top of the red giant branch in the colour-magnitude diagram of the globular cluster (Fig. \ref{fig:cmd_NGC6541}). From the light curve obtained it was possible to measure an amplitude of $A_{\mathrm{i}^{\prime}+\mathrm{z}^{\prime}}$=0.17 mag. We classified this star as a long-period irregular variable as it was not possible to find a period.

\begin{figure*}[htp!]
\centering
\includegraphics[scale=1 ]{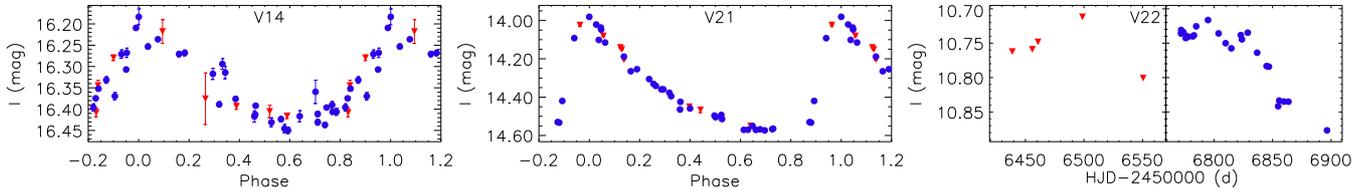}
\caption{NGC~6541: Light curves of the known and new variables discovered in this globular cluster. Symbols are the same as in Fig. \ref{fig:lc_chart1_NGC104}.}
\label{fig:lc_chart1_NGC6541}
\end{figure*}

\subsection{\textbf{NGC~6656 / C1833-239 / M 22}}\label{sec:NGC6656}

This globular cluster was discovered by Abraham Ihle in 1665\footnote{http://messier.seds.org/m/m022.html}. It is in the constellation of Sagittarius at 3.2 kpc from the Sun and 4.9 kpc from the Galactic centre. The cluster has a metallicity of 
[Fe/H]=-1.70 dex, a distance modulus of (m-M)$_V$=13.60 mag and the horizontal branch level is at V$_{HB}$=14.15 mag.

\begin{table*}[htp!]
\caption{NGC~6656: Ephemerides and main characteristics of the variable stars in the field of this globular cluster. Columns are the same as in Tab. \ref{tab:NGC104_ephemerides}.}
\label{tab:NGC6656_ephemerides}
\centering
\begin{tabular}{ccccccccc}
\hline\hline
Var id &   RA  &  Dec  &Epoch & $P$ & $I_{median}$ & $A_{\mathrm{i}^{\prime}+\mathrm{z}^{\prime}}$ & $N$ & Type\\
       & J2000 & J2000 & HJD  & d &      mag     &                mag                            &   &     \\
\hline
V35        & 18:36:24.051 & -23:54:29.53 & 2456472.8818 & 141(5) &  9.43 & 0.13 & 19 & SR \\
PK-06      & 18:36:25.228 & -23:54:37.44 & 2456789.8239 & 0.140851(4)  & 17.08 & 0.65 & 51 & EW \\
CV1        & 18:36:24.696 & -23:54:35.60 & 2456877.6355 &    --        & 17.87 & 3.28 & 51 & CV(DN) \\

\hline
\end{tabular}
\end{table*}

\subsubsection{Known variables}

There are of the order of 100 variable sources known for this globular cluster. Only 4 of these are inside the field of view of the reference image (V35, PK-06, PK-08 and CV1). PK-08 is too faint for us to extract a light curve.

\textbf{V35:} This star is the brightest star in our field of view and it is classified as a semi-regular variable. The star is at the top of the red giant branch in the colour-magnitude diagram shown in 
Fig. \ref{fig:cmd_NGC6656}. \cite{sahay14+02} found a period of $\sim$56 d. However, we found a period of $141\pm5$ d.

\textbf{PK-06:} This star is classified as an EW eclipsing variable. It was discovered by \cite{pietrukowicz03+01} and it is their star M22\_06. They found a period P=0.239431 d. However, it does not 
produce a good phased light curve in our data. In the analysis of this variable we found a period of P=0.140851 d which produces a better phased light curve (See Fig. \ref{fig:lc_chart1_NGC6656}). However, our 
phased light curve is still not as clear as that of \cite{pietrukowicz03+01}.

\textbf{CV1:} The variability of this star was first detected by \cite{sahu01+06} as a suspected microlensing event, but it was not until \cite{anderson03+02} that it was 
classified as a dwarf nova outburst. The analysis of our light curve for this source shows that it undergoes an outburst of $\sim$3 mag around HJD $\sim$2456877 which 
decays over $\sim$20d. It is in agreement with previous studies. \cite{anderson03+02} found that an earlier outburst lasted $\sim$20-26d with an amplitude of $\sim$3 mag 
peaking at $I\approx$15 mag. \cite{alonso-Garcia15+06} also observed the 2014 outburst seen in our data and reported a Ks-band brightening of 1 mag. In Fig. \ref{fig:lc_chart1_NGC6656}, we have 
plotted the light curve for this star.

\begin{figure}[htp!]
\centering
\includegraphics[scale=0.17]{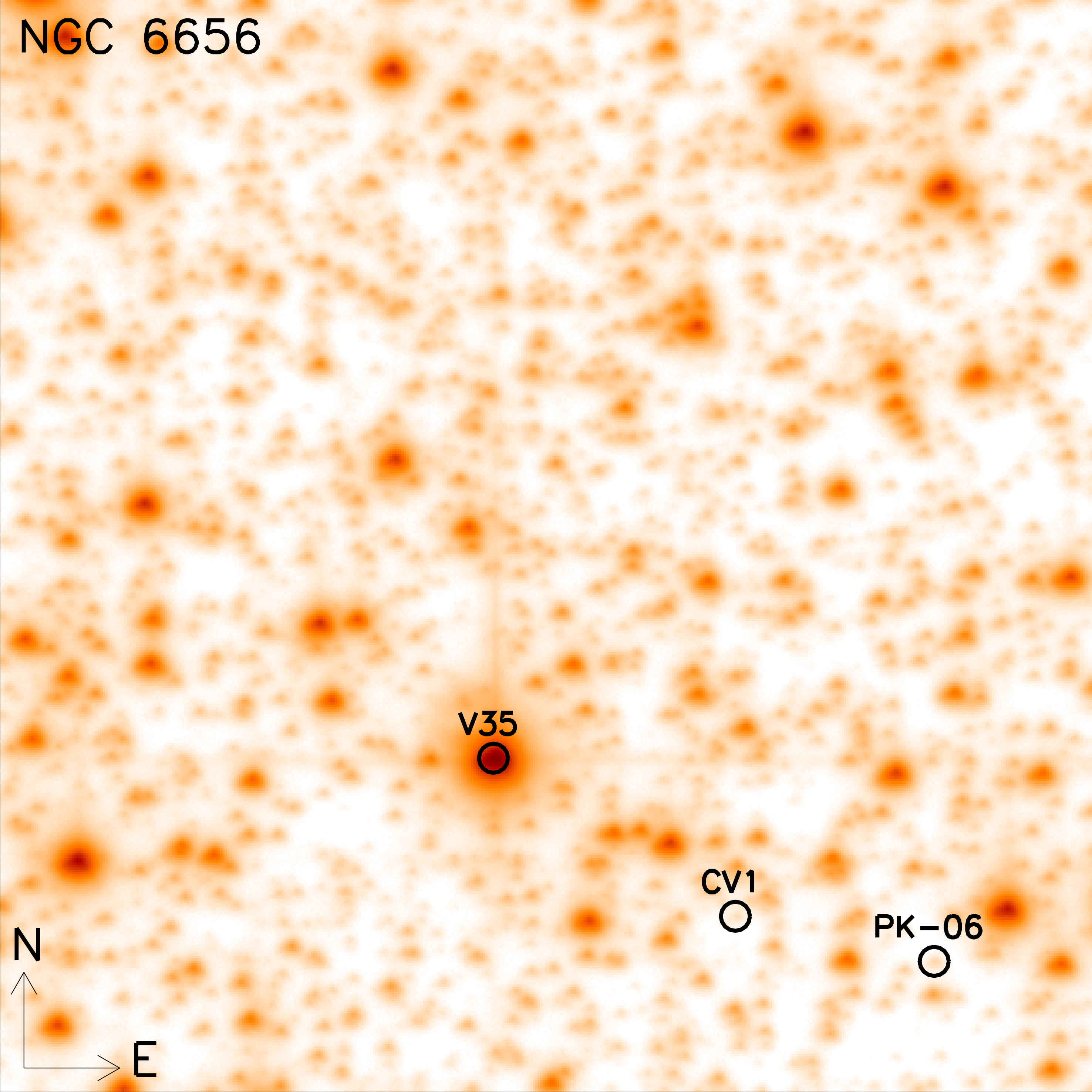}
\caption{Finding chart for the globular cluster NGC~6656. The image used corresponds to the reference image constructed during the reduction. All known variables and new discoveries are labelled. Image size is $\sim41\times41$ arcsec$^2$.}
\label{fig:finding_chart_NGC6656}
\end{figure}

Our search for new variable sources in this cluster did not yield any.

\begin{figure*}[ht]
\centering
\includegraphics[scale=1]{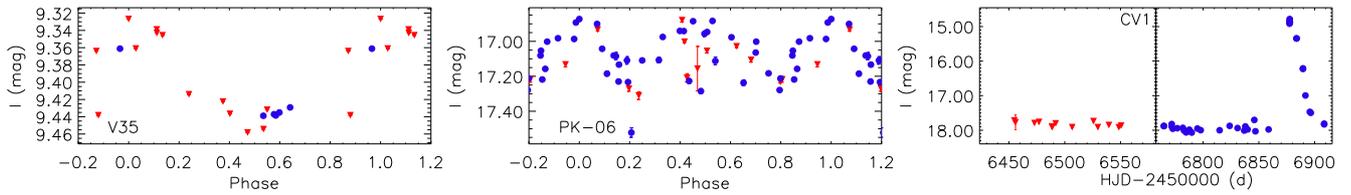}
  \caption{NGC~6656: Light curves of the known variables found in this globular cluster. Symbols are the same as in Fig. \ref{fig:lc_chart1_NGC104}.}
\label{fig:lc_chart1_NGC6656}
\end{figure*}

\begin{figure}[ht]
\centering
\includegraphics[scale=1]{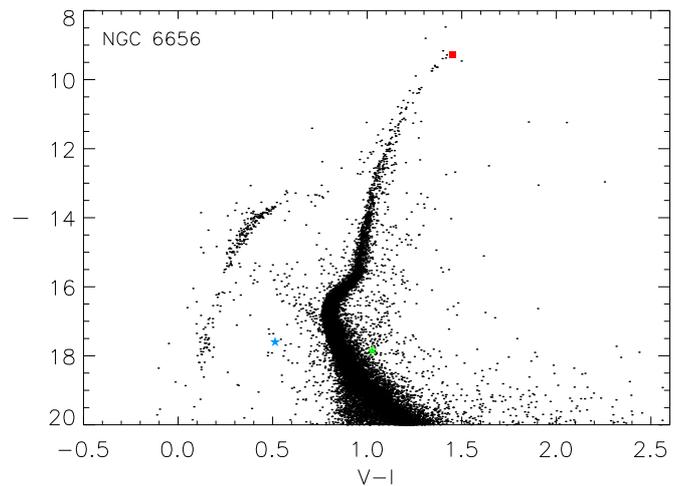}
\caption{Colour magnitude diagram for the globular cluster NGC~6656 built with V and I magnitudes available in the ACS globular cluster survey extracted from HST images. The variable stars are plotted in colour following the convention 
adopted in Tab. \ref{tab:var_type}. }
\label{fig:cmd_NGC6656}
\end{figure}

\subsection{\textbf{NGC~6681 / C1840-323 / M70}}\label{sec:NGC6681}

This globular cluster was discovered by Charles Messier in 1780\footnote{http://messier.seds.org/m/m070.html}. The 
globular cluster is in the constellation of Sagittarius at 9.0 kpc from the Sun and 2.2 kpc from the Galactic centre. It has a metallicity of 
[Fe/H]=-1.62 dex, a distance modulus of (m-M)$_{V}$=14.99 mag and the horizontal branch level is at V$_{HB}$=15.55 mag.

\begin{table*}[htp!]
\caption{NGC~6681: Ephemerides and main characteristics of one variable star in the field of this globular cluster. Columns are the same as in Tab. \ref{tab:NGC104_ephemerides}.}
\label{tab:NGC6681_ephemerides}
\centering
\begin{tabular}{ccccccccc}
\hline\hline
Var id &   RA  &  Dec  &Epoch & $P$ & $I_{median}$ & $A_{\mathrm{i}^{\prime}+\mathrm{z}^{\prime}}$ & $N$ & Type\\
       & J2000 & J2000 & HJD  & d &      mag     &                mag                            &   &     \\
\hline
V6 & 18:43:12.015 & -32:17:29.70 & 2456814.8334 & 0.341644(25) & 15.09 & 0.23 & 50 & RR1 \\

\hline
\end{tabular}
\end{table*}

\begin{figure}[ht]
\centering
\includegraphics[scale=1.0]{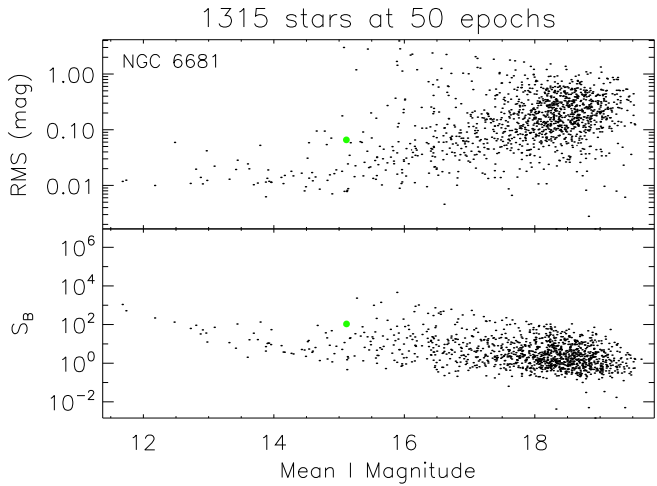}
\caption{Root mean square (RMS) magnitude deviation (top) and $S_B$ statistic (bottom) versus the 
mean $I$ magnitude for the 1315 stars detected in the field of view of the reference image for 
NGC~6681. The coloured point follows the convention adopted in Tab. \ref{tab:var_type} to identify 
the types of variables found in the field of this globular cluster.}
\label{fig:rms_sb_NGC6681}
\end{figure}

There are a total of 1315 light curves available for this cluster to be analysed. Most of them have 50 epochs. The RMS and $S_B$ diagrams can be found in Fig. \ref{fig:rms_sb_NGC6681}. Variable stars studied in this work are plotted in colour.

\subsubsection{Known variables}

This globular cluster has only 5 known variables, none of which are in the field of view of our reference image.

\cite{kadla96+02} reported several RR Lyrae candidates based on the position of the stars in the instability strip of the horizontal branch but none of them matched the position of the only variable star found in the field of view covered by our reference image which is the new variable V6 explained in the next section.

\begin{figure}[ht]
\centering
\includegraphics[scale=1]{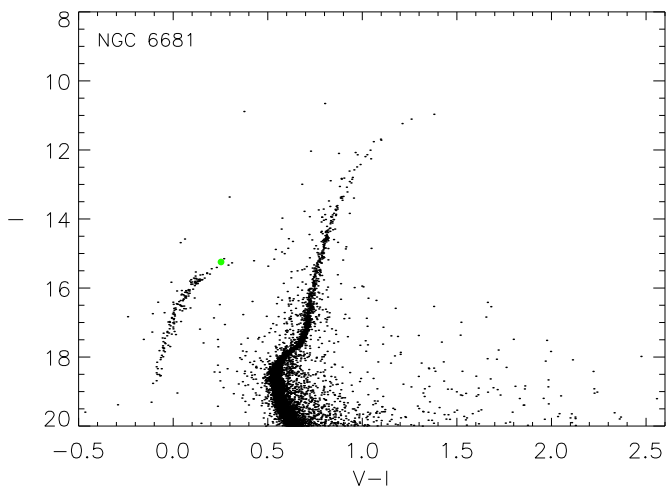}
\caption{Colour magnitude diagram for the globular cluster NGC~6681 built with V and I magnitudes available in the ACS globular cluster survey extracted from HST images. One variable star is plotted in colour following the convention 
adopted in Tab. \ref{tab:var_type}.}
\label{fig:cmd_NGC6681}
\end{figure}

\begin{figure}[ht]
\centering
\includegraphics[scale=0.17]{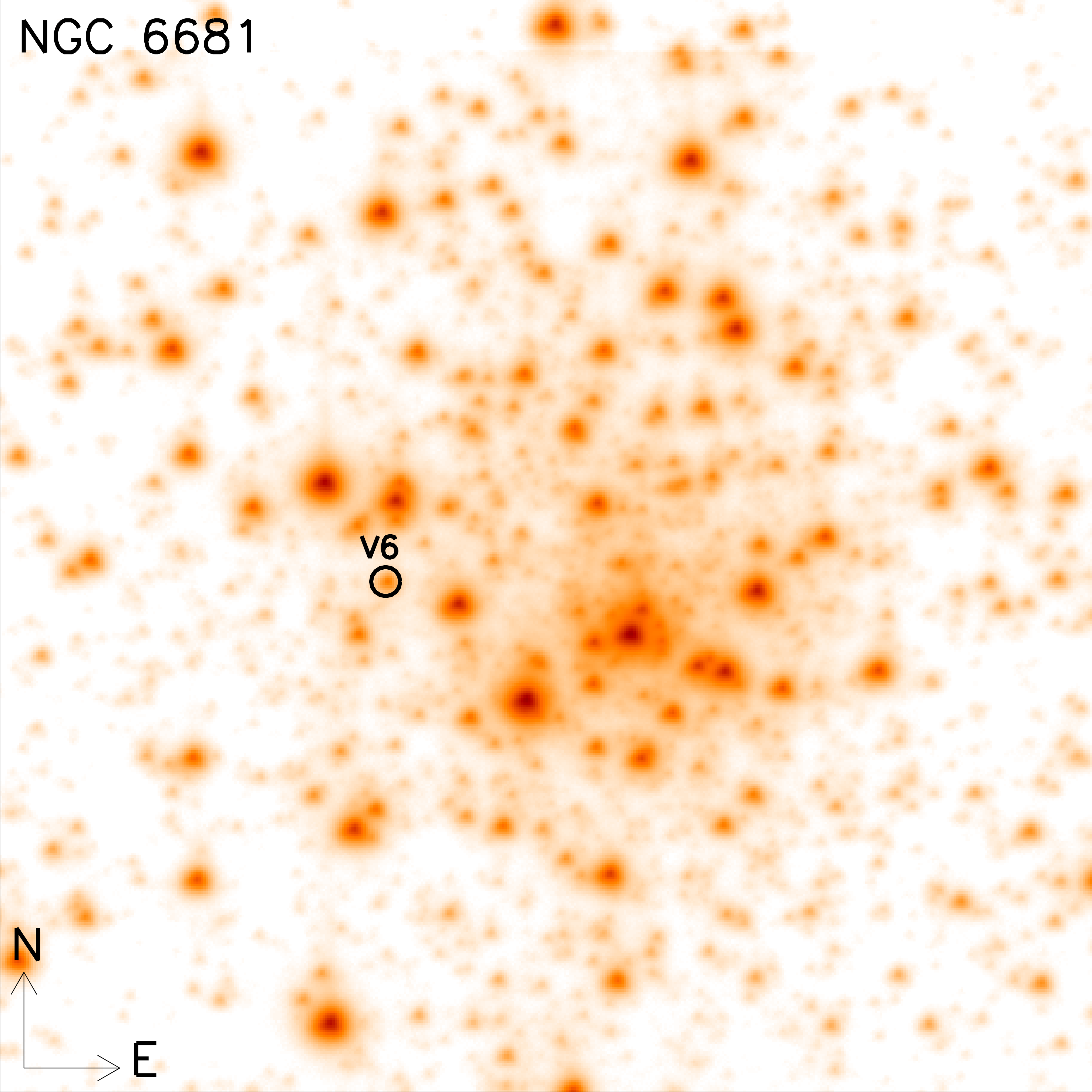}
\caption{Finding chart for the globular cluster NGC~6681. The image used corresponds to the reference image constructed during the reduction. All known variables and new discoveries are labelled. Image size is $\sim41\times41$ arcsec$^2$.}
\label{fig:finding_chart_NGC6681}
\end{figure}

\subsubsection{New variables}

One new RR Lyrae was discovered in this globular cluster.

\begin{figure}[htp!]
\centering
\includegraphics[scale=1.07]{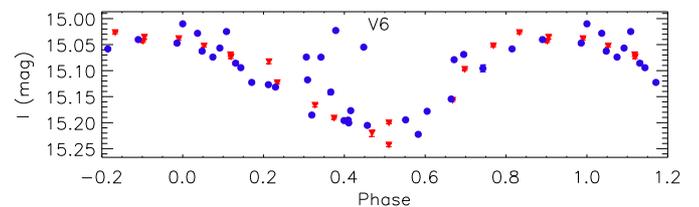}
  \caption{NGC~6681: Light curve of the new variable discovered in this globular cluster. Symbols are the same as in Fig. \ref{fig:lc_chart1_NGC104}.}
\label{fig:lc_chart1_NGC6681}
\end{figure}

\textbf{V6}: This star has an amplitude of 0.23 mag and a period of P=0.341644d. The star is placed just at the instability strip of the horizontal branch (see Fig. \ref{fig:cmd_NGC6681}). It 
is clearly a previously unknown RR Lyrae of type RR1. The light curve for this variable is shown in Fig. \ref{fig:lc_chart1_NGC6681}.

\subsection{\textbf{NGC~6723 / C1856-367}}\label{sec:NGC6723}

This globular cluster was discovered by James Dunlop in 1826\footnote{http://spider.seds.org/spider/MWGC/n6723.html}. 
It is in the constellation of Sagittarius at a distance of 8.7 kpc from the Sun and 2.6 kpc from the Galactic centre. It has a metallicity of [Fe/H]=-1.10 dex, a 
distance modulus of (m-M)$_V$=14.84 mag and the horizontal branch level is at V$_{HB}$=15.48 mag.

\begin{table*}[htp!]
\caption{NGC~6723: Ephemerides and main characteristics of the variable stars in the field of this globular cluster. Columns are the same as in Tab. \ref{tab:NGC104_ephemerides}.}
\label{tab:NGC6723_ephemerides}
\centering
\begin{tabular}{ccccccccc}
\hline\hline
Var id &   RA  &  Dec  &Epoch & $P$ & $I_{median}$ & $A_{\mathrm{i}^{\prime}+\mathrm{z}^{\prime}}$ & $N$ & Type\\
       & J2000 & J2000 & HJD  & d &      mag     &                mag                            &   &     \\
\hline
V8  & 18:59:34.678 & -36:37:42.33 & 2456908.5573 & 0.480278(49) & 15.08 & 0.71 & 16 & RR0 \\
V34 & 18:59:33.189 & -36:37:58.04 & 2456435.8962 & 0.531414(60) & 14.85 & 0.84 & 56 & RR0 \\
V35 & 18:59:32.963 & -36:38:01.46 & 2456524.8178 & 0.606451(78) & 14.80 & 0.32 & 56 & RR0 \\
V44 & 18:59:32.347 & -36:37:51.96 & 2456454.9399 & 0.440075(41) & 15.08 & 0.93 & 56 & RR0 \\

\hline
\end{tabular}
\end{table*}

\begin{figure}[ht]
\centering
\includegraphics[scale=1.0]{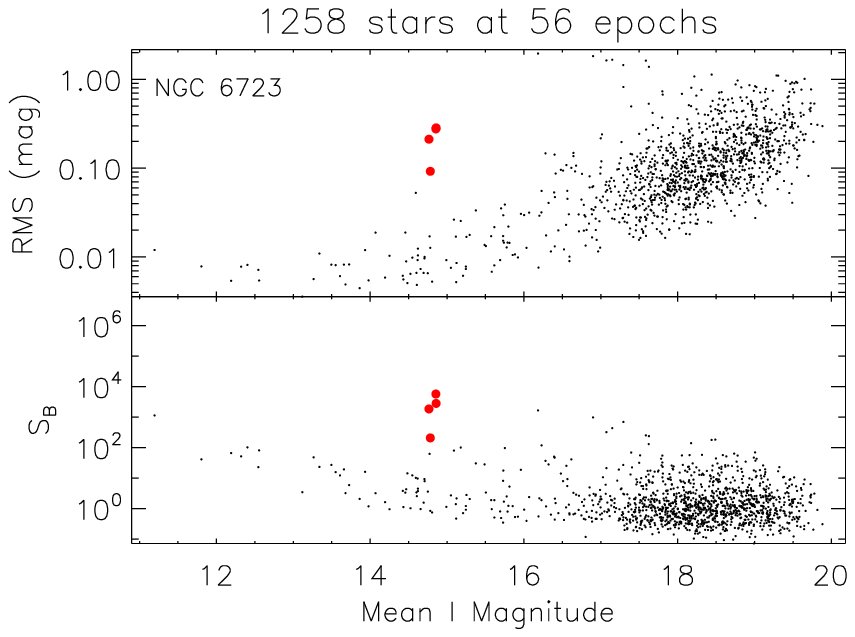}
\caption{Root mean square (RMS) magnitude deviation (top) and $S_B$ statistic (bottom) versus the 
mean $I$ magnitude for the 1258 stars detected in the field of view of the reference image for 
NGC~6723. Coloured points follow the convention adopted in Tab. \ref{tab:var_type} to identify 
the types of variables found in the field of this globular cluster.}
\label{fig:rms_sb_NGC6723}
\end{figure}

In Fig. \ref{fig:rms_sb_NGC6723}, the RMS and $S_B$ statistic diagrams for 1258 stars in the globular cluster NGC~6723 
are shown. Most of the light curves have 56 epochs. Variable stars analysed in this work are plotted in colour following the 
convention adopted in Tab. \ref{tab:var_type}.

\subsubsection{Known variables}

This globular cluster has 47 known variables stars in the Catalogue of Variable Stars in Galactic Globular Clusters 
\citep{clement01}. 43 are classified as RR Lyrae, two as semi-regular variables, one as a T Tauri star that does not 
seem to be a cluster member and one as a SX Phoenicis. Four of the known RR Lyrae are in the field of our reference 
frame (V8, V34, V35 and V44). They are RR0 pulsating stars. The star V8 was discovered by \cite{bailey1902} and although it is close 
to the reference image border, it was possible to obtain 16 data points. V34, V35 and V44 were discovered by \cite{lee14+05}.

In the colour magnitude diagram for this globular cluster (see Fig. \ref{fig:cmd_NGC6723}) these variables are placed just in the 
instability strip of the horizontal branch. Light curves for these variables are shown in Fig. \ref{fig:lc_chart1_NGC6723}. 

\begin{figure}[ht]
\centering
\includegraphics[scale=1]{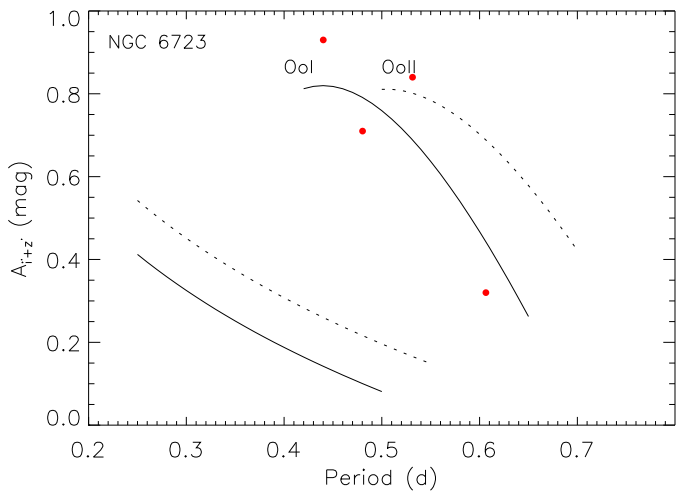}
\caption{Period-amplitude diagram for the globular cluster NGC~6723. The previously known RR Lyrae are plotted.}
\label{fig:pad_NGC6723}
\end{figure}

\begin{figure}[ht]
\centering
\includegraphics[scale=1]{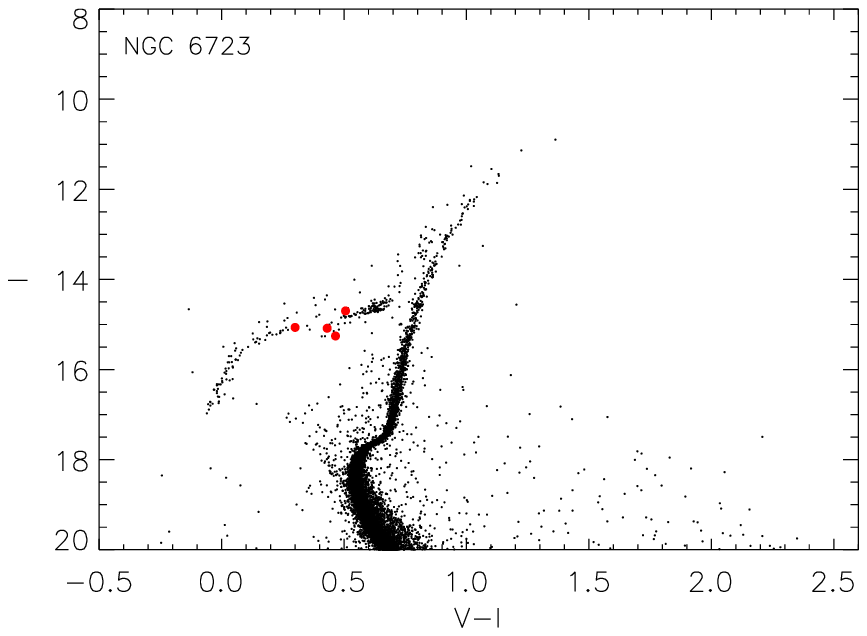}
\caption{Colour magnitude diagram for the globular cluster NGC~6723 built with V and I magnitudes available in the ACS globular cluster survey extracted from HST images. The variable stars are plotted in colour following the convention 
adopted in Tab. \ref{tab:var_type}.}
\label{fig:cmd_NGC6723}
\end{figure}

The four RR Lyrae in the field of our images are plotted in the period-amplitude diagram shown in Fig. \ref{fig:pad_NGC6723}. 
It is possible to see that even though V34 and V35 have the Blazhko effect, all of them are following the model of fundamental 
mode pulsating stars for Oosterhoff type I which is consistent with the Oosterhoff classification for this cluster \cite[see][and references therein]{lee14+05,kovacs86+02}.

The periods that we derive for the 4 RR Lyrae stars are perfectly consistent with those derived by \cite{lee14+05} using a 10 
year baseline. Furthermore, the Blazhko effect in V34 and V35 is also evident in our light curves.

\begin{figure}[ht]
\centering
\includegraphics[scale=0.17]{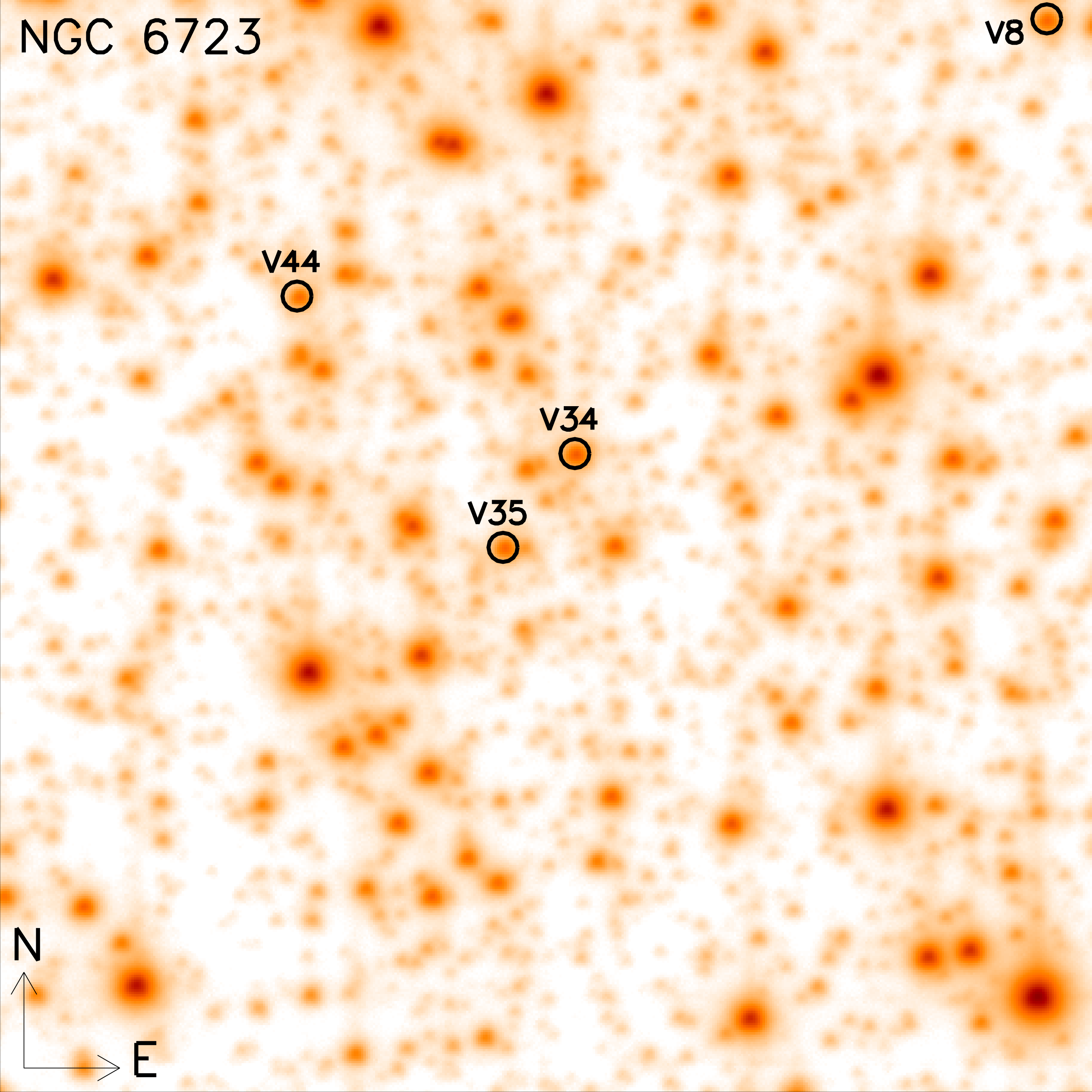}
\caption{Finding chart for the globular cluster NGC~6723. The image used corresponds to the reference image constructed during the reduction. All known variables and new discoveries are labelled. Image size is $\sim41\times41$ arcsec$^2$.}
\label{fig:finding_chart_NGC6723}
\end{figure}

No new variable stars were found in the field covered by the reference image for this globular cluster.

\begin{figure*}[htp!]
\centering
\includegraphics[scale=1]{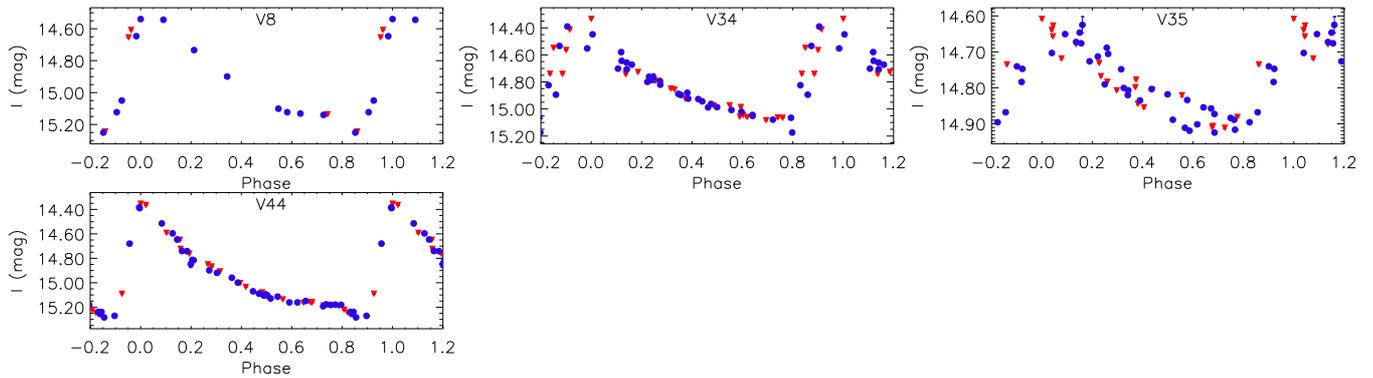}
  \caption{NGC~6723: Light curves of the known variables in this globular cluster. Symbols are the same as in Fig. \ref{fig:lc_chart1_NGC104}.}
\label{fig:lc_chart1_NGC6723}
\end{figure*}

\subsection{\textbf{NGC~6752 / C1906-600 / C93}}\label{sec:NGC6752}

This globular cluster was discovered by James Dunlop in 1826\footnote{http://spider.seds.org/spider/MWGC/n6752.html}. It is in 
the constellation of Pavo with a distance of 4.0 kpc from the Sun and 5.2 kpc from the Galactic centre. It has a metallicity of 
[Fe/H]=-1.54 dex, a distance  modulus of (m-M)$_V$=13.13 mag and the level of its horizontal branch is at V$_{HB}$=13.70 mag.

\begin{table*}[htp!]
\caption{NGC~6752: Ephemerides and main characteristics of one variable star in the field of this globular cluster. Columns are the same as in Tab. \ref{tab:NGC104_ephemerides}.}
\label{tab:NGC6752_ephemerides}
\centering
\begin{tabular}{ccccccccc}
\hline\hline
Var id &   RA  &  Dec  &Epoch & $P$ & $I_{median}$ & $A_{\mathrm{i}^{\prime}+\mathrm{z}^{\prime}}$ & $N$ & Type\\
       & J2000 & J2000 & HJD  & d &      mag     &                mag                            &   &     \\
\hline
V26 & 19:10:51.494 & -59:58:56.67 & -- & -- & -- & -- & 78 & CV(DN) \\

\hline
\end{tabular}
\end{table*}

\subsubsection{Known variables}

There are 32 known variable sources in this globular cluster which are listed in the Catalogue of Variable Stars in Galactic 
Globular Clusters \citep{clement01}. Six of them are in the field of view of our reference image. Three are millisecond 
pulsars (PSRB, PSRD and PSRE) discovered by \cite{damico01+04} and \cite{damico02+06}. We could not find an optical counterpart 
in our reference image. Three more are Dwarf Novae (V25, V26, V27) discovered by \cite{thomson12+08} of which V26 is the only object in which we could detect variability.

\begin{figure}[htp!]
\centering
\includegraphics[scale=1.07]{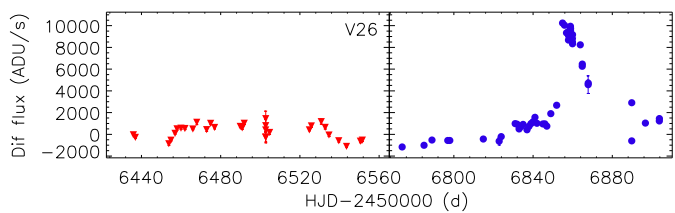}
  \caption{NGC~6752: Light curve of the variable V26 in this globular cluster. Symbols are the same as in Fig. \ref{fig:lc_chart1_NGC104}. We plot the quantity $f_{\mathrm{diff}}(t)/p(t)$ since a reference flux 
is not available.}
\label{fig:lc_chart1_NGC6752}
\end{figure}

\textbf{V26:} This star is classified as a Dwarf Nova. A finding chart for V26 is already available in \cite{thomson12+08}. It was not possible 
to detect this star in the reference image in its quiescent state. However, 
it was possible to clearly detect one outburst in the difference images. Due to this, in Fig. \ref{fig:lc_chart1_NGC6752} the 
difference fluxes measured for this star are plotted against Heliocentric Julian Day. It is possible to see that in the 2013 
campaign the star was in its quiescent state. In 2014 observing season, the outburst started around HJD 2456840 with a maximum at HJD $\sim$2456858.8532 and lasted $\sim$80 d.

We did not find any evidence in our data for new variable stars in this globular cluster.

\section{Conclusions}\label{sec:conclusion}

   \begin{enumerate}
      \item The central regions of 10 globular clusters were studied.
      \item The benefits of using EMCCDs and the shift-and-add technique were shown.
      \item A total of 12541 stars in the fields covered in the center of each globular cluster were studied for variable star detection.
      \item Light curves for 31 previously known variables are presented (3 L, 2 SR, 20 RR Lyrae, 1 SX Phe, 3 cataclysmic variables, 1 EW and 1 NC). Period improvement 
      and ephemerides are presented for most of them as well.
      \item The discovery of 30 variable stars is presented (16 L, 7 SR, 4 RR Lyrae, 1 SX Phe and 2 NC). Period and ephemerides calculations were also done.
      \item Light curves are available for all variable stars studied in this work.
      \end{enumerate}

\begin{acknowledgements}
Our thanks go to Christine Clement for clarifying the known variable star content of each cluster and the numbering systems within each one while we 
were working on these clusters. This support to the astronomical community is very much appreciated. The Danish 1.54m telescope is operated based on 
a grant from the \emph{Danish Natural Science Foundation (FNU)}. RFJ thanks Ian Taylor for sorting out all my computational needs on my arrival at St Andrews. RFJ also thanks Katherine White for her comments and suggestions in the improvement of my English language. This publication was made possible by NPRP grant \# X-019-1-006 from the \emph{Qatar National Research Fund (a member of Qatar Foundation)}. The statements made herein are solely the responsibility of the authors. KH acknowledges 
support from STFC grant ST/M001296/1. GD acknowledges Regione Campania for support from POR-FSE Campania 2014-2020. TH is supported by a Sapere Aude 
Starting Grant from the Danish Council for Independent Research. Research at Centre for Star and Planet Formation is funded by the Danish National 
Research Foundation. TCH acknowledges support from the Korea Research Council of Fundamental Science \& Technology (KRCF) via the KRCF Young Scientist 
Research Fellowship Programme and for financial support from KASI travel grant number 2013-9-400-00 \& 2014-1-400-06. OW and J. Surdej acknowledge 
support from the Communaut\'{e} fran\c{c}aise de Belgique - Actions de recherche concert\'{e}es - Acad\'{e}mie Wallonie-Europe. This work has made 
extensive use of the \emph{ADS} and \emph{SIMBAD} services, for which we are thankful.
\end{acknowledgements}

\bibliographystyle{biblio}

\end{document}